\documentclass[a4paper,twoside, 12pt]{article}
\usepackage{graphicx}
\usepackage{fullpage}
\usepackage{epsfig}
\usepackage{fancyhdr}
\usepackage{fancybox}
\usepackage{pst-all}

\usepackage{amssymb}
\usepackage{wasysym}

\usepackage{lscape,amsfonts,graphicx,fancyhdr}

\usepackage{verbatim}

\catcode`\@=11
\def\@citex[#1]#2{\if@filesw\immediate\write\@auxout{\string\citation{#2}}\fi
  \def\@citea{}\@cite{\@for\@citeb:=#2\do
    {\@citea\def\@citea{,\penalty\@m}\@ifundefined
       {b@\@citeb}{{\bf ?}\@warning
       {Citation `\@citeb' on page \thepage \space undefined}}%
\hbox{\csname b@\@citeb\endcsname}}}{#1}}

\def\citer{\@ifnextchar [{\@tempswatrue\@citexr}{\@tempswafalse\@citexr[]}}
\def\@citexr[#1]#2{\if@filesw\immediate\write\@auxout{\string\citation{#2}}\fi
  \def\@citea{}\@cite{\@for\@citeb:=#2\do
    {\@citea\def\@citea{--\penalty\@m}\@ifundefined
       {b@\@citeb}{{\bf ?}\@warning
       {Citation `\@citeb' on page \thepage \space undefined}}%
\hbox{\csname b@\@citeb\endcsname}}}{#1}}
\catcode`\@=12

\begin{document}
\setlength{\parskip}{0ex}
\begin{center}
  {\Large \bf A New Measure of Fine Tuning}\\[1cm]
{\large Peter Athron and D.J.~Miller \\ [1cm]
 {\it Department of Physics and Astronomy, University of Glasgow,} \\
{\it Glasgow G12 8QQ, United Kingdom \\}}
\end{center}
%%\classification{<PACS numbers;\texttt{http://www.aip.org/pacs/index.html}>}
%%\keywords      {<Enter Keywords here>}

\begin{abstract}
\noindent
The solution to fine tuning is one of the principal motivations for
Beyond the Standard Model (BSM) Studies. However constraints on new
physics indicate that many of these BSM models are also fine tuned
(although to a much lesser extent). To compare these BSM models it is
essential that we have a reliable, quantitative measure of tuning. We
review the measures of tuning used in the literature and
propose an alternative measure. We apply this measure to several toy
models and the Minimal Supersymmetric Standard Model.

\end{abstract}

\section{Introduction}

Fine tuning appears in many areas of particle physics and cosmology,
such as the Standard Model (SM) Hierarchy Problem and the Cosmological
Constant Problem. These problems imply that the the universe we live in is a
very atypical scenario of the theories we use to describe it.  The
contortion required to reproduce observation makes such theories seem
unnatural, motivating many studies of Beyond the Standard Model
(BSM) physics.%

However many of the models constructed to solve fine tuning, also
exhibit some degree of tuning themselves. In the absence of data,
while we await the LHC, naturalness is used to compare models and
judge their viability. Great importance has been attached to small
differences in the levels of tuning when comparing models, so it is
important that naturalness and fine tuning are rigorously understood
and measured accurately.

For example the Hierarchy Problem is one of the fundamental
motivations of low energy supersymmetry (SUSY) (for a review see
Ref.\cite{Martin:1997ns}).  If the SM is an effective theory, valid
up to the Planck scale, then the inclusion of supersymmetric partners
for every SM particle leads to the cancellation of quadratic
divergences in the loop corrections to the Higgs mass.  This removes
the need for fine tuning of ${\cal O}(10^{34})$ between the tree-level
mass parameter and the Planck Mass, allowing the Higgs boson to be
naturally light.

Unfortunately current limits on superpartner masses may imply fine
tuning in the most studied model, the Minimal Supersymmetric Standard
Model (MSSM). The minimisation of the Higgs potential sets the square
of the $Z$ boson mass, $M_Z^2$, in terms of the supersymmetry breaking
scales. In the MSSM the tree-level expression for this is,
\begin{equation} \label{Mzsqpred}
M_Z^2 = \frac{2
(m^2_{H_d} - m^2_{H_u}\tan^2{\beta})}{\tan^2{\beta}-1 } - 2|\mu|^2
  \rm{,}
\end{equation}
where $\tan{\beta}$ is the ratio of vacuum expectation values, $\mu$
the bilinear Higgs superpotential parameter, and $m_{H_u}$ and
$m_{H_d}$ are the up and down type Higgs scalar masses respectively.

Lower bounds on the masses of the supersymmetric particles and the
Higgs translate to lower bounds on the parameters appearing on the
right hand side of Eq.\ (\ref{Mzsqpred}). If, for example, one of the
parameters is $1\,$TeV, then to cancel this contribution and
give $M_Z = 91.1876 \pm 0.0021\,$GeV~\cite{Yao:2006px}, another parameter (or combination
of parameters) would have to be tuned to the order of one part in a
hundred.

Including loop corrections to Eq.\ (\ref{Mzsqpred}) and examining the
experimental constraints, one finds that the largest term is from
corrections involving the heaviest stop. This can be written as
\cite{Chang:2005ht},
\begin{equation}
\delta m^2_{H_u} = -\frac{3y_t^2}{8 \pi^2}(m^2_{\tilde{t}_l} + m^2_{\tilde{t}_r})\log\left(\frac{\Lambda}{m_{\tilde{t}}}\right)\rm{,} 
\end{equation}
where $\Lambda$ is the high scale at which the soft stop masses,
$m_{\tilde{t}_l}$ and $m_{\tilde{t}_r}$, are generated from the
supersymmetry breaking mechanism and $y_t$ is the top Yukawa
coupling. A heavy physical stop mass ($m_{\tilde{t}} \gtrsim 500$ GeV)
is needed to provide radiative corrections to the light CP even Higgs
mass, $m_{h^0}$ of the form,
\begin{equation}   
\delta m^2_{h^0} = \frac{3 v^2
y_t^2}{4\pi^2}\sin^4{\beta}
\log\left(\frac{m_{\tilde{t}_l}m_{\tilde{t}_r}}{m_t^2}\right)\rm{,}
\end{equation}
which are large enough to evade the LEP constraints on it's mass
($\geq 114$ GeV). So the Little Hierarchy Problem is really about the
tension between the masses of the $Z$ boson, the heaviest stop squark
and the light Higgs.

The desire to solve this ``Little Hierarchy Problem'' has motivated a
flood of activity in the construction of supersymmetric models
\citer{Choi:2005uz,Chang:2006ra}. There
is also increased interest in studying alternative solutions to the SM
Hierarchy problem \citer{Casas:2005ev,Chacko:2005un}. In
addition to ensuring such models satisfy phenomenological constraints
it is essential that the naturalness is examined using a reliable,
quantitative measure of tuning.  \\

\noindent In Ref.\cite{Barbieri:1987fn} Barbieri and Giudice use a
measure of tuning, originally proposed in Ref.\cite{Ellis:1986yg}, for an
observable, $O$, with respect to a parameter, $p_i$,
\begin{equation}\label{Trad_meas}
\triangle_{BG}(p_i) = \Big{|}\frac{p_i}{O(p_i)}\frac{\partial O(p_i)}{\partial p_i}\Big{|} \rm{.}
\end{equation}
A large value of $\triangle_{BG}(p_i)$ implies that a small change in
the parameter results in a large change in the observable, so the
parameters must be carefully ``tuned'' to the observed value. Since
there is one $\triangle_{BG}(p_i)$ per parameter, they define the
largest of these values to be the tuning for that point,
\begin{equation}\label{max_eq}
\triangle_{BG} = \rm{max}(\{\triangle_{BG}(p_i)\})\rm{.}
\end{equation}
They then make the aesthetic choice that a tuning, $\triangle_{BG} >
10$ is fine tuned.

This measure has been used extensively in the literature to quantify
tuning in the MSSM
\citer{deCarlos:1993yy,Kobayashi:2006fh}
and to examine tuning in other models and theories
e.g.~\cite{Chang:2005ht},\citer{Dermisek:2005ar,Dermisek:2006py}.
However other measures have
also been proposed and used in the literature.

Motivated by global sensitivity, which will be discussed in the next
section, Anderson and Castano
\citer{Anderson:1994dz,Anderson:1996ew}
propose that tuning should be measured with,
\begin{equation}
\triangle_{AC}(p_i) = \frac{\triangle_{BG}(p_i)}{\bar{\triangle}_{BG}(p_i)}\rm{,}
\end{equation}
where they choose the ``average'' sensitivity,
$\bar{\triangle}_{BG}(p_i)$, not to be the mean, but instead defined
by,
\begin{equation} 
\bar{\triangle}_{BG}^{-1}(p) = \frac{\int pf(p)\triangle_{BG}^{-1}(p)dp}{pf(p)\int dp}\rm{.}
\end{equation}
where f(p) is the probability distribution of parameter
$p$. Individual $\triangle_{AC}(p_i)$ are combined in the same manner
as the individual $\triangle_{BG}(p_i)$ were,
\begin{equation} 
\triangle_{AC} =  \rm{ max} (\{\triangle_{AC}(p_i) \} )\rm{.} 
\end{equation}
 
There is some dispute within the literature as to whether or not Eq.\
(\ref{max_eq}) is the best way of choosing a final tuning value from
the set $\{\triangle_{BG}(p_i)\}$. In
\cite{Casas:2005ev},\citer{Casas:2003jx,Casas:2006bd}
the individual $\triangle_{BG}(p_i)$ are be combined as if
uncorrelated,
\begin{equation}\label{Uncorrelated_Errors}
\triangle_E = \sqrt{\displaystyle\sum_i\triangle_{BG}^2(p_i)}\rm{.} 
\end{equation}

Several other measures have been proposed
\citer{Ciafaloni:1996zh,Giusti:1998gz}, but
will not be discussed here.  \\

\noindent In Section \ref{Literature} we detail some limitations of the
traditional measure of tuning, $\triangle_{BG}$, used for this Little
Hierarchy Problem. We then describe our fundamental notion of tuning,
and how this principle can be applied to construct quantitative
measures of tuning in Section \ref{Constructing_tuning_measure}. This
leads us to present a new tuning measure in
Section \ref{A_New_Measure}, which is also a generalisation of the
traditional measure that overcomes the limitations outlined earlier.
This measure is applied to several toy models in
Section \ref{Toy_Models} to demonstrate how it works and compare the
results it produces with those from other measures. Finally in
Section \ref{MSSM_Tuning} we apply our measure to the Little
Hierarchy Problem for a selection of Minimal SUperGRAvity (MSUGRA)
inspired points.

\section{Limitations of the Traditional Measure}
\label{Literature}
Despite the wide use of $\triangle_{BG}$ it has several limitations
which may obscure the true picture of tuning:
\begin{itemize}\setlength{\parskip}{0ex}
\item variations in each parameter are considered separately;
\item only one observable is considered in the tuning measure, but there may be tunings in several observables;
\item it does not take account of global sensitivity;
\item only infinitesimal variations in the parameters are considered;
\item there is an implicit assumption that the parameters come from uniform probability distributions.
\end{itemize}
\noindent  Tuning is really concerned with how the parameters
are combined to produce an unnatural result. If one measures tunings
for each parameter individually, there is no clear guide how to
combine these tunings to quantify how unnatural this cancellation
is. This has led to two alternative approaches in the literature,
Eq.\ (\ref{max_eq}) and Eq.\ (\ref{Uncorrelated_Errors}); the only way
to determine if either $\triangle_{BG}$ or $\triangle_E$ combines
sensitivities correctly is to compare them with a generalisation of
$\triangle_{BG}(p_i)$ that varies all of the parameters
simultaneously.

Secondly, some theories may contain significant tunings in more than
one observable. We want to know how can these tunings be combined to
provide a single measure. For example it is reported in
Refs.\cite{Chankowski:1998za,Ellis:2002rp}, and more recently in
Refs.\cite{King:2006tf,King:2006cu}, that the MSSM also requires
tuning in the relic density of the dark matter ($\rho$). To measure
the tuning for some particular set, $S' =\{M_Z',\rho'\}$, of these
observables we should determine how atypical predictions like $S'$ are
in the theory.  There are four classes of scenario which are
significant: the first where both $M_Z$ and $\rho$ are similar to
their value in $S'$; two more classes where only one of $M_Z$ or
$\rho$ is similar to it's value in $S'$; one with neither observable
similar to $S'$. Tunings in these two observables should be combined
in a manner which measures how atypical scenarios in the first class
are, without double counting scenarios which appear in the final
class. Only a tuning measure which considers the observables
simultaneously can achieve this.
 
A third problem, first mentioned by Anderson and Castano
\cite{Anderson:1994dz} is that the traditional measure picks up
global sensitivity as well as true tuning.  $\triangle_{BG}$ is
really a measure of sensitivity. Consider the simple mapping $f: x
\rightarrow x^n $, where $n \gg 1$. Applying the traditional measure
to $f(x)$ gives $\triangle_{BG} = \triangle_{BG}(x) = n$. Since
$\triangle_{BG}$ is independent of $x$, we follow the example of
\cite{Anderson:1994dz} and term this {\em global sensitivity}. Since
$\triangle_{BG}(x_1) - \triangle_{BG}(x_2) = 0$ for all $x_1, x_2$,
there is no {\em relative sensitivity} between points in the parameter
space.

If we use $\triangle_{BG}$ as our tuning measure then $f(x)$ appears
fine tuned throughout the entire parameter space. This contrasts with
our fundamental notion of tuning being a measure of how atypical a
scenario is.  A true measure of tuning should only be greater than
one when there is relative sensitivity between different points in
the parameter space.

Another concern is that $\triangle_{BG}$ only considers infinitesimal
variations in the parameters. Since MSSM observables are complicated
functions of many parameters, it is reasonable to expect some
complicated distribution of the observables about that parameter
space. There may be locations where some observables are stable
(unstable) locally, but unstable (stable) over finite variations.

Finally, there is also an implicit assumption that all values of the
parameters in the effective softly broken Lagrangian ${\cal L}_{SUSY}$
are equally likely. However they have been written down in ignorance
of the high-scale theory, and may not match the parameters in, for
example, the Grand Unified Theory (GUT) Lagrangian, ${\cal
L}_{GUT}$. Any non-trivial relation between these different sets of
parameters may alleviate or exacerbate the fine tuning problem.

While some of the alternative measures in the literature are
motivated by one of these issues, no proposed measure fully addresses
all of them.  

\section{Constructing Tuning Measures}
\label{Constructing_tuning_measure}
A physical theory is fine tuned when generic scenarios of the theory
predict very different physics to that which is observed. For the theory
to agree with observation the parameters must be adjusted very
carefully to lie in an extremely narrow range of values. Insisting
that the physics described by the theory is similar to that observed,
shrinks the acceptable volume of parameter space. When in this tiny
volume even small adjustments to the parameters will dramatically
change the physics predicted, so fine tuning may also be characterised
by instability. It is this instability which the traditional measure
is exploiting.

Instead we wish to construct a tuning measure which determines how
rare or atypical certain physical scenarios are. The most direct way
to do this is to compare the volume of parameter space, $G$,
that is \emph{similar} to some given scenario with the \emph{typical}
volume, $T$, of parameter space formed by scenarios which are
\emph{similar} to each other.

If all the parameters $\{p_i\}$ are drawn from a uniform probability
distribution then the probability of obtaining a scenario in $G$ is,
$G/V$, where $V$ is the volume of parameter space formed by
all possible parameter choices.  Similarly $T/V$ gives the
probability of obtaining a scenario in volume $T$. We may then define
tuning as $\hat{\triangle} = T/G$, to quantify the
relative improbability of scenarios \emph{similar} to our given
scenario in comparison to the \emph{typical} probability.

\begin{figure}[ht]
\begin{center}
\includegraphics[width=70mm,clip=true]{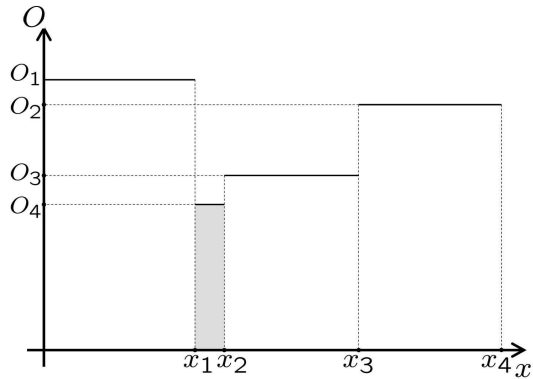}
\end{center}
\caption{A toy example with an observable, $O$, which depends on a parameter $x$. }
\label{next_simple_tuning}
\end{figure}

To place this within a quantitative framework we must define what we
mean by ``similar'' and ``typical''. This will be dealt with
later. First, though, consider the toy example presented in
Fig.~\ref{next_simple_tuning},
showing an observable, $O$, which depends on a parameter,~$x$.  Here
there are four clearly distinct groups of observable scenario
\mbox{$(O = O_1,O_2,O_3,O_4)$} and ``similar'' can be replaced with
equal. Given one of these groups of scenarios, $O=O_i$, the volume $G$
is the length (one dimensional volume) of parameter space with
$O=O_i$.  For example, for $O_4$ we have $G = x_2 -x_1$. Next we must
define our ``typical'' volume, $T$ formed by these distinct groups of
scenarios.  In this simple example an obvious choice is to define $T$
as the mean volume (length) of parameter space formed by scenarios in
the same group.  So $T = \frac{1}{4} (x_1 + (x_2 - x_1) + (x_3-x_2) +
(x_4 -x_3)) = \frac{x_4}{4}$. The tuning required to get $O=O_4$ is
then $\hat{\triangle} = \frac{x_4}{4(x_2- x_1)}$, which conforms to
our intuitive expectation.

In more realistic examples the definitions of ``similar'' and
``typical'' will not be so trivial. The definitions must be chosen to
fit the type of problem one is considering.  In the simple example
given above the problem was that scenarios where $O=O_4$ occupied a
smaller proportion of the parameter space than other values, $O =
O_1,O_2,O_3$.

In hierarchy problems the concern is that one (or more) observable
is much smaller than another observable, despite depending on 
common parameters.  The requirement that one observable is large
forces the theory into a region of parameter space where generic
points also predict a large value for the second observable(s).

So ``similar'' must be related to the size of the observables.  For
example, one might consider ``similar'' to observable $O_i'$ to mean
observables ``of the same order''as $O_i'$. A sensible definition of
$G$ is then the volume of parameter space where $\frac{1}{10} \leq
\frac{O_i}{O_i'} \leq 10$, for all observables $O_i$.  However it is
not clear that this is more appropriate than some other choice such as
$\frac{1}{2} \leq \frac{O_i}{O_i'} \leq 2$. So generally $G$ can be
defined by a class of parameter space volumes formed from
dimensionless variations in the observables $a \leq \frac{O_i}{O_i'}
\leq b$. Different values of $a$ and $b$ quantify different
definitions of ``similar'' and are therefore different fine-tuning
questions. In comparison, the one dimensional measure $\Delta_{BG}$ is
a ratio of infinitesimal lengths, so implicitly adopts the choice
$a,\,b \to 1$. One can imagine cases where this would be a bad choice
(e.g.\ an observable which oscillates quickly when the parameter is
varied), so care must be taken to choose $a$ and $b$ sensibly (i.e.\
ask the correct question).

When a large hierarchy between observables requires a large
cancellation between parameters, as in the traditional hierarchy
problem, the region of parameter space which can provide the correct
observables (the volume $G$) is much smaller than one would expect
(i.e.\ it is ``fine-tuned''). We must compare this volume with the
``typical volume'' of parameter space, $T$, that one would expect if
no fine-tuning were present. The remaining question is then, how do we
define this ``typical volume''?

One might suggest that this typical volume should be the average of
volumes $G$ throughout the whole parameter space, $\langle G
\rangle$. However, the measure would then depend only on how far
parameters are from some hypothesised upper limits on their
values. For example, an observable $O$ which depends on a parameter
$p$ according to $O = \alpha p$ will display fine-tuning for small
values of $p$ if one chooses the maximum possible value of $p$ to be
large, even though there is no cancellation present. This is not the
`fine-tuning' we are trying to probe; we want to gain insight into the
unnatural cancellation between parameters, so $T$ must be anchored to the
specific parameter point to be tested.
 
 We can do this by adopting the same notion of ``similar'' that we
used to define $G$. We introduce a volume $F$ which is formed from
dimensionless variations $[a,b]$ in the parameters. A comparison of
$F/G$ at different points in the parameter space, provides a test of
whether $G$'s variation is due to a simple scaling with the parameters
(as described above for $O = \alpha p$), or due to some ``unnatural''
effect such as fine-tuning. Consequently one should compare $F/G$ with
its average value over the entire space, $\langle F/G
\rangle$. Reverting to our previous terminology, the ``typical''
volume which one would have expected to form from dimensionless
variations in the parameters about $\{p_i'\}$, is
\begin{equation}
T = \frac{F(\{p_i'\})}{\langle\frac{F}{G}\rangle}\rm{.}
\end{equation} 

\section{A New Measure}
\label{A_New_Measure}
Following the above discussion and motivated by the limitations of the
traditional measure, we propose a new measure of tuning.

\begin{figure}[ht]
\begin{center}
    \includegraphics[height=30mm]{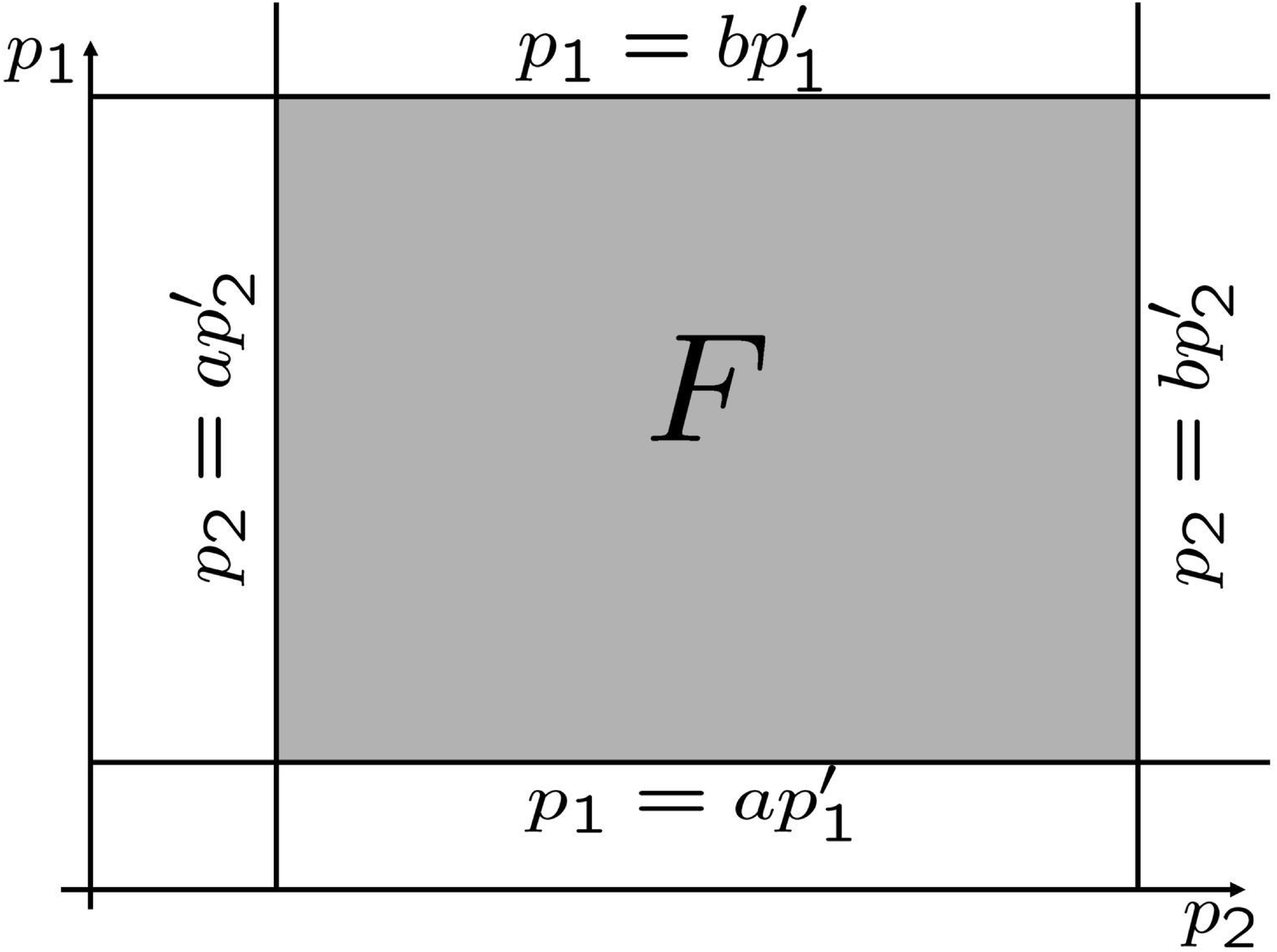}  \hspace*{8mm}
    \includegraphics[height=30mm]{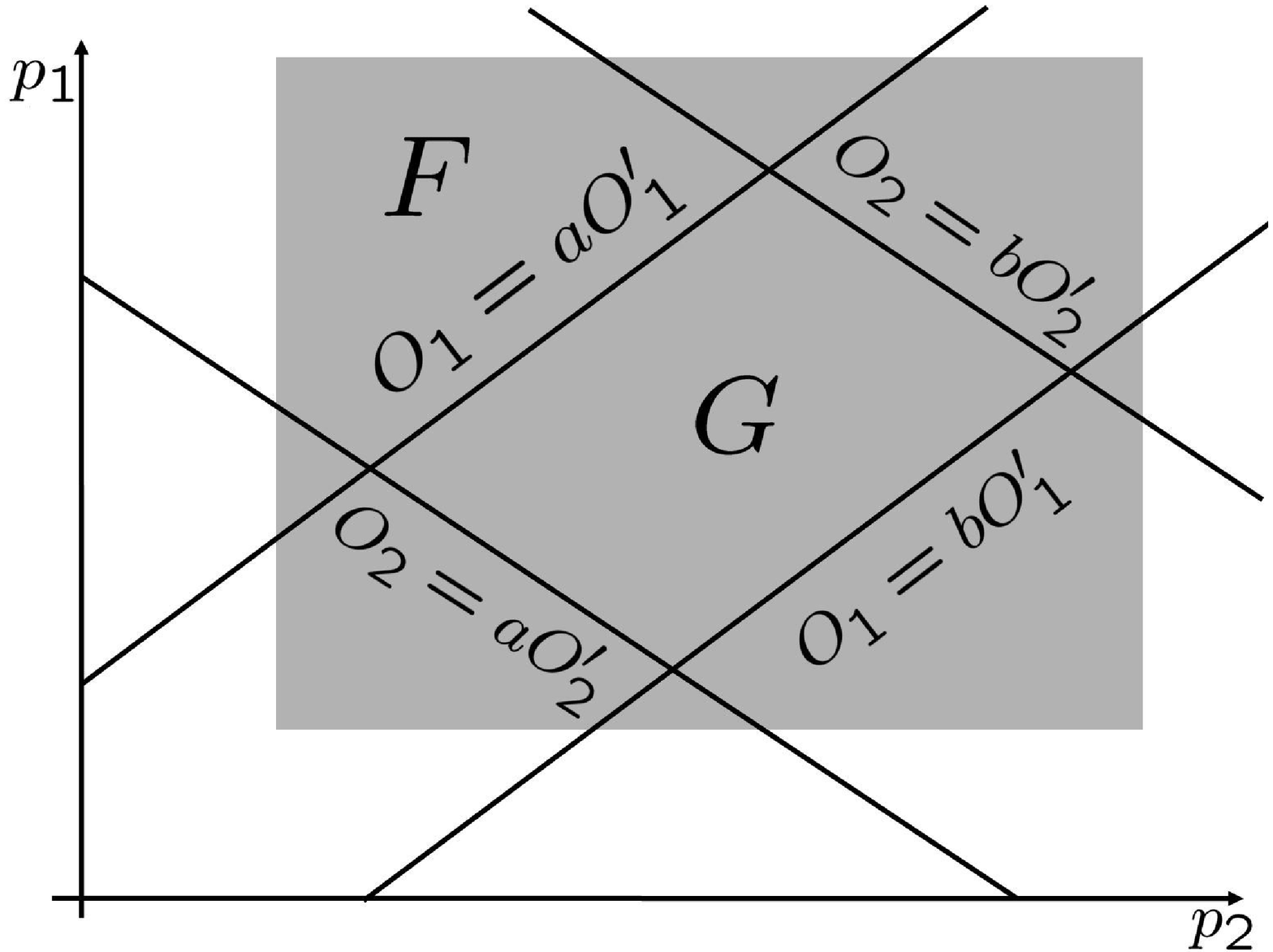} \hspace*{8mm}
    \includegraphics[height=30mm]{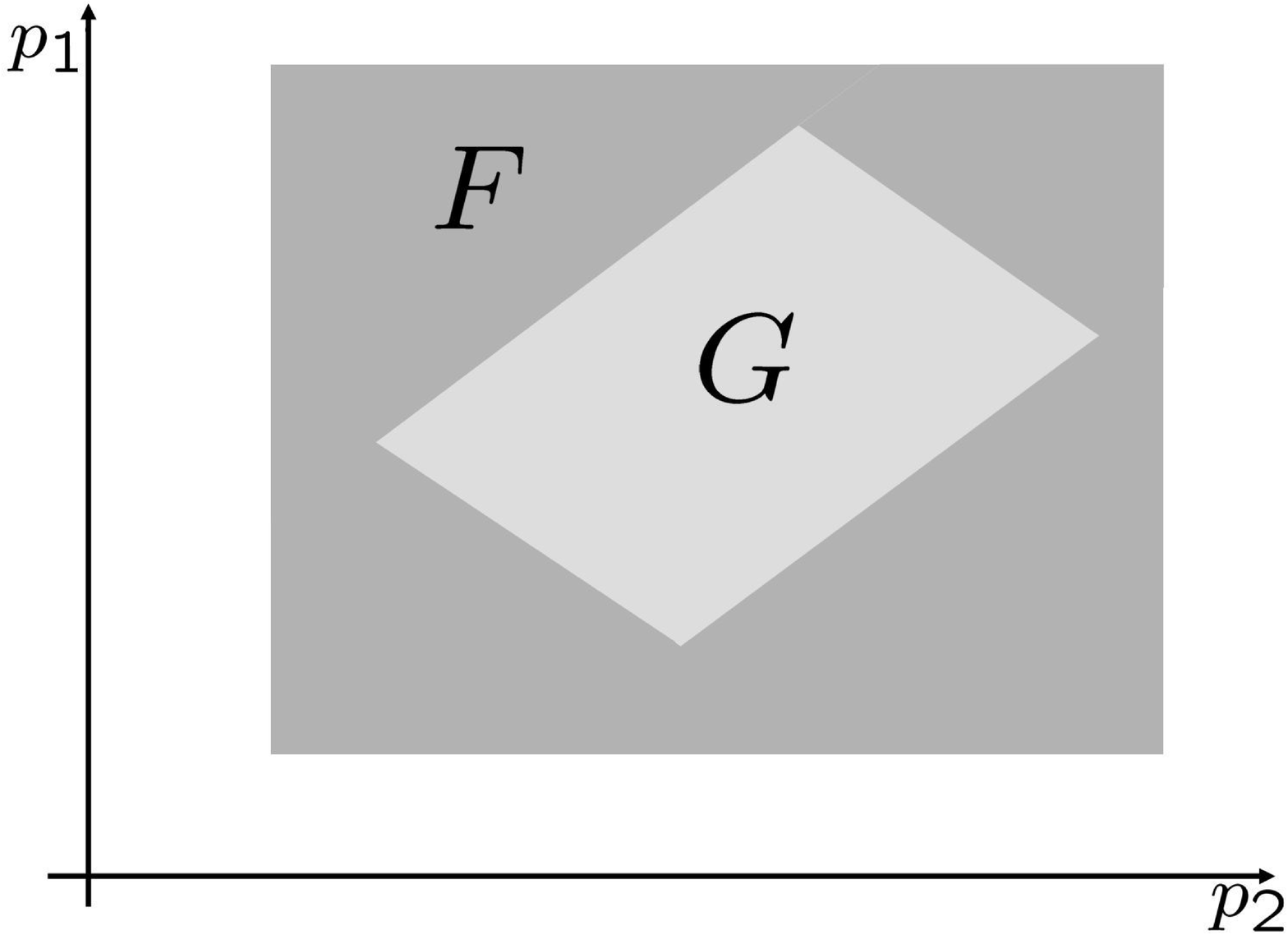}
\end{center}
\caption{{\it Left}: In two dimensions the bounds placed on the
parameters, $a\leq\frac{p_i}{p_i'}\leq b$, appear as four lines in
parameter space giving the dark grey area (2d volume), $F$. {\it
Middle}: Bounds on the two observables,
$a\leq\frac{O_j(\{p_i\})}{O_j(\{p_i'\})}\leq b$ introduce four more
lines giving the volume $G$. {\it Right}: Two dimensional volumes
(areas) $F$ (dark grey) and $G$ (light grey). }
\label{Tuning_illustration}
\end{figure}

We define two volumes in parameter space for every point $P'
\{p_i'\}$.  Let $F$ be the volume of dimensionless variations in the
parameters over some arbitrary range $[a,b]$, about point $P'$, i.e.\
the volume formed by imposing \mbox{$a\leq\frac{p_i}{p_i'}\leq
b$}. Similarly let $G$ be the volume in which dimensionless variations
of the observables fall into the same range $[a,b]$, i.e.\ the volume
constrained by $a\leq\frac{O_j(\{p_i\})}{O_j(\{p_i'\})}\leq b$.
Volumes $F$ and $G$ are illustrated for a two dimensional example in
Fig.~\ref{Tuning_illustration}.

We define an unnormalised measure of tuning with, 
\begin{equation} \triangle = \frac{F}{G} \rm{.}\end{equation}
This is sufficient for comparing different regions of parameter space
within a given model as the normalisation factor will be common.  To
compare tuning in different models we need to include normalisation,
\begin{equation} 
\hat{\triangle} = \frac{1}{\bar{\triangle}}\frac{F}{G} \rm{,}
\end{equation}
with,
\begin{equation} 
\bar{\triangle} = \left\langle\frac{F}{G}\right\rangle 
= \frac{\int dp_1 ... dp_n 
\frac{F}{G}(\{p_i\}, \{O_i\})}{\int dp_1 ... dp_n}\rm{.}
\end{equation}
Notice that this measure does not depend on experimental
constraints. In naturalness problems such constraints should only rule
out the point, $P'$, around which we make variations to test fine
tuning. If $P'$ is not experimentally excluded, we should not impose
experimental constraints on nearby points $\{P_i\}$ used to probe fine
tuning. Fine tuning quantifies how unnatural a region of parameter
space is and this is a feature of the theory, not our experimental
knowledge.

$\hat{\triangle}$ quantifies the restriction on parameter space. This
is more in touch with our intuitive notion of tuning than the
stability of the observable.  Notice that with only one or two
parameters and no global sensitivity, $\triangle_{BG}$ also describes
restriction of parameter space and yields the same results as our new
measure. However it is important to recognise that $\triangle_{BG}$'s
ability to do this leads to its utility as a tuning measure
there. Equally its failure to do so in many dimensions demonstrates
its limitation.
\\

\noindent Consider fine tuning for a single observable which depends
on more than one parameter, Even though the true tuning for any
physical scenario should be described using all available observables,
it is often useful to define individual tunings for each observable
separately. However, in this case, the volume $G$ is unbounded, since
a single observable can only constrain one combination of parameters.

To resolve this difficulty one must either reduce the number of
parameters to one or introduce some other bounds on $G$.  The former
reintroduces the problem of combining tunings for individual
parameters and a better procedure is to restrict $G$ to be within $F$.
Here we are trying to pick up how much of the restriction in parameter
space is due to this particular observable.  The assumption is made
that if all other observables were natural then they would restrict
$G$ no more than $F$ does.  Therefore we define $G_{O_j}$ to be the
volume restricted by $a\leq\frac{O_j(\{p_i\})}{O_j(\{p_i'\})}\leq b$
and $a\leq\frac{p_i}{p_i'}\leq b$. Tuning is then defined by,
\begin{equation} 
\hat{\triangle}_{O_j} 
= \frac{1}{\left\langle\frac{F}{G_{O_j}}\right\rangle}\frac{F}{G_{O_j}} \rm{,}
\end{equation}
This definition is applied to obtain individual tunings in the MSSM in
Section \ref{MSSM_Tuning}.  \\

\noindent Like $\triangle_{BG}$ and $\triangle_{AC}$, $\triangle$
depends upon the choice of parameterisation. Since tuning is about the
restriction of the parameter space this seems unavoidable. To examine
different choices of parametrisation one must redefine volumes $F$ and
$G$ in terms of the new parameters and normalise the tuning by taking
the average over the new parameter space.  \\

\noindent Since much of the motivation behind developing this measure
was to generalise $\triangle_{BG}$ so that many parameters and many
observables are considered simultaneously, it is interesting to look
at how the two measures are related.

Consider a theory with one observable, $y$ which has a linear
dependence on a single parameter, $x$, with the value of that
parameter being drawn from a uniform probability distribution.  At the
parameter point $(x_0,y_0)$, notice that, $\triangle_{BG} = |x_o/y_o
\partial y / \partial x|$, while we can see $F = (b-a)x_o$ and $G
=\frac{\partial x}{\partial y} (b-a)y_o$, so,
\begin{equation}
\triangle =\frac{F}{G} =  
\frac{b x_o - ax_o}{b y_o -a y_o}\frac{\partial y}{\partial x} 
= \triangle_{BG}. 
\end{equation}

\noindent Similarly Anderson and Castano's measure may be written as,
\begin{equation}
\triangle_{AC} 
= \frac{x_o}{y_o}\frac{\partial y}{\partial x} \frac{ \int dx' y(x') 
\frac{\partial x'}{\partial y(x')}}{x_o \int dx'} 
= \frac{\int dx' y(x')}{y_o \int dx'}
\end{equation}
Now notice that $ \langle G \rangle = \frac{\partial x}{\partial y}
\frac{\int dx' (b-a)y(x')}{\int dx'}$, so,
\begin{equation}\triangle_{AC} = \frac{\langle G \rangle}{G}. \end{equation}

In Section \ref{Constructing_tuning_measure}, we pointed out the
difficulty in using $\langle G \rangle/G$ as a tuning measure
and this will be further illustrated in Section \ref{Toy_Models} when
we look at results for our measure and $\triangle_{AC}$ for a toy
version of the SM Hierarchy problem.

\section{Toy Models}
\label{Toy_Models}
We now compare some of the tuning measures for various toy models and
discuss the implications. In each of these examples we will assume a
uniform probability distribution for the parameters.
 
Table \ref{Toy_tuning_table} compares the analytical results of
various tuning measures for the simple models with only one parameter
and one observable.  With only one parameter it is trivially the case
that $\triangle_E = \triangle_{BG}$, so it is not included. \\

\setlength{\tabcolsep}{0.5mm}
\begin{table}
\begin{center}
\begin{tabular}{l||ccccc} 
\hline 
& \footnotesize $\triangle_{BG}$ 
& \footnotesize $\triangle$ 
& \footnotesize $\triangle_{AC}$ 
& \footnotesize $\hat{\triangle}$\\ \hline\\ 
  \footnotesize Toy SM   
& \footnotesize $1+\frac{C\Lambda^2}{m_H^2}$ 
& \footnotesize $1+\frac{C\Lambda^2}{m_H^2}$ 
& \footnotesize $ \frac{m_{H max}^2 + m_{H min}^2 }{2 m_H^2}  $ 
& \footnotesize $ \frac{m_0^2}{m_H^2 + \frac{m_H^2 C 
                  \Lambda^2}{m_{0 max}^2 - m_{0 min}^2 } 
                  \ln{ \frac{ m_{0 max}^2 - C \Lambda^2}{  m_{0 min}^2 - C \Lambda^2}}}   $ \\ \\%%\hline
  \footnotesize $f(x) = x^n$       
& \footnotesize $n$ 
& \footnotesize $\frac{b-a}{b^{1/n}- a^{1/n}}$  
& \footnotesize $ \frac{x_{max} + x_{min}}{2x} $ 
& \footnotesize $ 1 $\\ \\ %%\hline 
  \footnotesize $g(x) = e^{kx}$      
& \footnotesize $|kx|$  
& \footnotesize $ \frac{(b-a)|kx|}{ln{\frac{b}{a}}}$  
& \footnotesize $1$  & $ \frac{2 x}{x_{min}+ x_{max}}  $\\  \\[0.5ex]%%\hline 
  \footnotesize Proton Mass~    
& \footnotesize $2\frac{8 \pi^2}{b_3 g_3^2} $  
&  \footnotesize $\frac{(b-a)}{(\frac{-k}{g_3^2\ln{b} -k })^{1/2} - (\frac{-k}{g_3^2\ln{a} -k })^{1/2}}\;\;\;$  
& \footnotesize $\frac{(g_{max} + g_{min})(g^2_{max} + g^2_{min})}{4 g_3^3}$ &$\approx \frac{g_{max} g_{min}}{g_3^2}$\\ \\\hline 
\end{tabular}
\caption{Tuning measures for models with only one parameter and one observable}
\label{Toy_tuning_table}
%%\label{sps1a_point}
\end{center}
\end{table}

In the first row of Table \ref{Toy_tuning_table} are the
results for a toy version of the Standard Model Hierarchy Problem,
where we know the tuning is enormous.  Here there is only one
observable, the physical Higgs mass, $m_H$. At one loop we write,

\begin{equation}\label{SM_Hierarchy} m^2_H = m^2_0 - C\Lambda^2, \end{equation}
and treat only the bare mass squared, $m^2_0$ as a parameter.
$\Lambda$, the Ultra-Violet cutoff, is taken to be the Planck Mass or
some other fixed scale, while  $C$ is a positive constant.

Our measure was obtained by simply varying the tree-level mass
parameter over the arbitrary range $[am_0^2,bm_0^2]$ and applying the
same dimensionless variations to the observable. This gives $F
=(b-a)m_0^2$ and $G = (b-a)m_H^2$, leading to the result for
$\triangle$ shown. Notice that the arbitrary range $[a,b]$ has fallen
out of the result and it matches that obtained using the traditional
measure, as shown earlier for all linear functions.

We also determine $\hat{\triangle}$, and $\triangle_{AC}$. In both
cases this introduces a dependence on the allowed range of $m_0^2$ in
the theory, so we specify $m_{0 min}^2 \leq m_0^2 \leq m_{0 max}^2$,
and present results where $m_{0 min}^2 > C \Lambda^2$, though similar
results can be obtained for other scenarios. These bounds give the
total allowed range of the parameter in this model and should not be
confused with the range of dimensionless variations which appears in
the definition of $F$. If we take the range of variation to be large,
\mbox{$m_{0 max}^2 - m_{0 min}^2 \gg C \Lambda^2$}, then \mbox{$
\hat{\triangle} \approx \frac{m_0^2}{m_H^2} = \triangle_{BG}$}.
Alternatively, if we choose a very narrow range of variation about $C
\Lambda^2 + \mu_H^2$, where $\mu_H \approx 100 \,\rm{GeV}$, then
$\hat{\triangle}$ is very small. 

This is intuitively reasonable.  Imagine some compelling theoretical
reason for the bare mass to be constrained close to the cutoff,
e.g.~$C\Lambda^2 \leq m_0^2 \leq C \Lambda^2 + (150\,\rm{GeV})^2$. In
light of this, the case for new physics at low energies would be
dramatically weakened.  Indeed it is precisely because there is no
such compelling reason that we worry about the hierarchy problem and
look to BSM physics such as supersymmetry to explain how we can have
$m_H \ll M_{\rm Planck}$.

Now let us compare this with the result for $\triangle_{AC}$. $m_{H
{\rm min}}$ and $m_{H {\rm max}}$ are the extremum values of $m_H$,
dictated by the extremum values of $m_0$.  Notice that as \mbox{$m_{0
max}^2 \to m_{0 min}^2$} we have $m_{H {\rm min}} \to m_H$ and $m_{H
{\rm max}} \to m_H$, so $\triangle_{AC} \rightarrow 1$.  However a
fundamental difference between our measure and $\triangle_{AC}$ is
that the latter will give a large tuning for any $m_H^2 \ll
\frac{1}{2}(m_{0 max}^2 + m_{0 min}^2) - C \Lambda^2$.  If the upper
bound is chosen such that, $m_{0 max}^2 \gg m_0^2$, then even a Higgs
mass of ${\cal O}(m_0^2)$ will appear fine tuned.  This measure is not
sensitive to the unnatural cancellation which causes our concern.
Instead it is sensitive to the fact that large values of $m_H^2$ take
up a much larger volume of parameter space than small values of
$m_H^2$.  This would be true even if the Higgs mass was described by
$m_H^2 = m_0^2$, with no unnatural cancellation.\\

Also shown in Table \ref{Toy_tuning_table} are the results for the
simple functions $f(x) = x^n$ and $g(x) = e^{kx}$. Earlier we showed
there was no relative sensitivity in $f(x)$. While $\triangle_{BG}$
and $\triangle$ can be large despite the absence of relative
sensitivity, our measure, $\hat{\triangle}$, is exactly unity for all
$x$. Anderson and Castano's measure does remove the global
sensitivity, but their tuning criterion prefers the observable to be
as large as possible.  For $g(x)$, while there is relative sensitivity
between different values of $x$, the constant factor $k$ makes
$\triangle_{BG} > 10 $ for all $|x| > 10/k$.  For $\triangle$
the situation is similar, with $\triangle > 10 $ for all $|x| >
10/K$, where $K=k(b-a)/\ln \frac{b}{a}$. In
$\hat{\triangle}$ the effect of $K$ is removed and though tuning still
increases with $x$, this is now contextualised by comparing it to
$\bar{\triangle}$. It is interesting that our measure considers $f(x)$
to have consistently no tuning ($\hat{\triangle} = 1$), whereas it is
for $g(x)$ that $\triangle_{AC} = 1$ for all $x$.  \\

The original illustration of global sensitivity presented by Anderson
and Castano in\ Ref.\cite{Anderson:1994dz} was for the proton
mass. The proton can be much lighter than the Planck Mass without fine
tuning because the renormalisation group equations (RGE) lead to only
a logarithmic dependence on high scale quantities.  However, by using
the one loop RGE for the QCD coupling, $\alpha_3$, and equating the
proton mass to the QCD scale\footnote{For details see
\cite{Anderson:1994dz}}
\begin{equation}\label{proton_mass} 
M_{\rm Proton} \sim \Lambda_{QCD} 
= C \exp{\left[-\frac{8 \pi^2}{b_3 g_3^2}\right]}\rm{,}
\end{equation}
where $g_3$ is the strong gauge coupling evaluated at the Planck
scale, $M_{\rm Planck}$, and C is a positive constant. As they
demonstrated, this gives $\triangle_{BG}(g_3) > 100$.

The analytical results for tuning in the mass of the proton, using
Eq.\ (\ref{proton_mass}), are shown in the final row of Table
\ref{Toy_tuning_table}.  Notice that while the unnormalised tunings
are both $ > 100$, $\triangle_{AC}$ and $\hat{\triangle}$ are small.
The latter has been determined only approximately in the limit where
$g^2 \ln{b} \ll k$ and $g^2 \ln{a} \ll k$ for all
$g_{min} \leq g_3 \leq g_{max}$, where $k = 8 \pi^2/b_3$.  \\

In these one parameter examples the need for a normalised tuning
measure is apparent. However $\triangle_{AC}$ diverges significantly
from our new measure, which in many of these simple one dimensional
models is equivalent to normalising the traditional measure with it's
mean value.

It is also interesting that even after accounting for global
sensitivity some of these one dimensional functions may still show
some small degree of tuning.  This opens up the possibility that
changing the parameterisation of the effective low energy theory might
exacerbate or alleviate the tuning problem. Finding choices of
parametrisation which reduce tuning could allow us to select high
scale theories which are preferential in terms of naturalness. This point
has not appeared in the literature and merits investigation.  However
we do not address this here but leave it for a future study.\\

Now we consider models with more than one parameter. In these cases
$\triangle_E$ diverges from $\triangle_{BG}$ and we must compare each
of these with $\triangle$.

First we return to the SM hierarchy problem, but this time treat $m_H$
as a function of two parameters, $m_0^2$ and $\Lambda^2$.  In the one
dimensional example the tension between the weakness of gravitation
(the large Planck Mass) and a light Higgs mass was examined indirectly
by choosing the Planck mass to be a fixed constant in theory.  We now
take a more direct route with two observables $m_H^2$ and
$M^2_{\rm Planck}$ (``observed'' to be large due to the weakness of
gravitation), predicted from the parameters with,
\begin{equation} 
M_{\rm Planck}^2 
=  \Lambda^2,\: \: \: \:\quad m^2_H = m^2_0 - C\Lambda^2 .
\end{equation}     

We are still predicting $m_H^2$ from Eq.~(\ref{SM_Hierarchy}) and have
not split up any of the terms to introduce new cancellations, so we
expect to simply reproduce the same result for $\triangle$ as we
obtained in the one parameter toy SM model.  However, the method
applied provides a simple illustration of how our measure works with
more than one parameter. We have a two dimensional parameter space, so
allowing the parameters to vary about some point $P'(m_0^2,
\Lambda^2)$ over the dimensionless interval $[a,b]$ defines an area,
$F$, in this space. Clearly the bounds from dimensionless variations
in $M_{\rm Planck}^2$ are the same as those from $\Lambda^2$, while the
bounds from dimensionless variations in $m_H^2$ introduce two new
lines in the parameter space.

\begin{figure}[ht]
\begin{center}
\includegraphics[height=50mm]{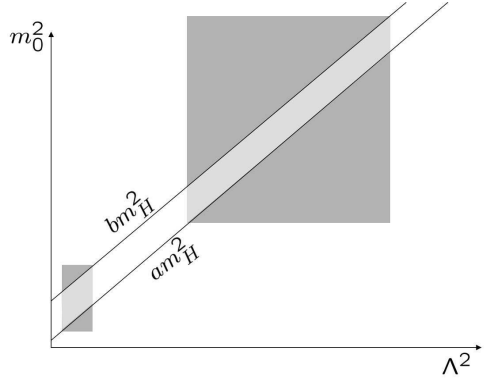}
\end{center}
\caption{The two dimensional volumes (areas) $F$ (dark grey) and $G$
(light grey) for two different points in the two dimensional parameter
space.}
\label{2d_SM}
\end{figure}

 This is shown in Fig.~\ref{2d_SM} for two different points. In the
first point, the values of the parameters are of the same order as the
observable, $m_H^2$, because we have chosen a small value of
$M_{\rm Planck}$. So $G$ is not much smaller than $F$.  For the other
point $M^2_{\rm Planck} \gg m_H^2$, resulting in an $F$ much larger than
$G$ and fine tuning.  Of course neither of these points are
representative of the weakness of gravitation we observe.  A point
with \mbox{$M_{\rm Planck} = 10^{19}\,{\rm GeV}$} and \mbox{$m_H =
120\, {\rm GeV}$}, would have $ F \gg G$ to such an extent that a
graphical illustration is not possible.

In general the areas are, $F = (b-a)^2m_0^2\Lambda^2$ and $G
=(b-a)^2\Lambda^2m_H^2$ so,
\begin{equation} 
\triangle = 1 + \frac{C\Lambda^2}{m_H^2}  = \triangle_{BG}.
\end{equation}
In this simple case we find the same result as the traditional
measure. Combining $\triangle_{BG}(\Lambda)$ and
$\triangle_{BG}(m_0^2)$ as if they are uncorrelated, gives,
\begin{equation} 
\triangle_E = \frac{\sqrt{C^2\Lambda^4 + m_0^4}}{m_H^2} {\rm .}
\end{equation}
With $C\Lambda^2$ and $m_0^2$ both $\gg m_H^2$, i.e.\ fine tuned
scenarios, this gives us $\triangle_E \approx
\sqrt{2}\triangle$. While our measure does not deviate from
$\triangle_{BG}$ in this simple example, models with additional
parameters allow the observable to be obtained from cancellation of
more than two terms, complicating the fine tuning picture.  \\

%%\subsection{Three Dimensional Toy Model} 
\noindent We now look at a model with four observables, $M^2$,
$M_1^2$, $M_2^2$, $M_3^2$, and three parameters, $p_1^2$, $p_2^2$,
$p_3^2$, described by,
\begin{equation}\label{3d_tuning} 
M^2 = c_1 p_1^2 - c_2 p_2^2 + c_3 p_3^2 {\rm .}
\end{equation}
\begin{equation} 
M_1^2   = p_1^2{\rm ,} \;\;\:\:  
M_2^2   = p_2^2{\rm ,}  \;\; \:\:   
M_3^2   = p_3^2          {\rm .}
\end{equation}
For a point $(m_1^2, m_2^2, m_3^2)$, in the three dimensional
parameter space, the traditional measure gives $\triangle_{BG}(p_i)=
c_im_i^2/M^2$ (no sum over $i$ is implied), so,
\begin{equation} 
\triangle_{BG} = {\rm max}\left\{ \frac{c_im_i^2}{M^2}\right\} \:\;\:\;\: 
{\rm and} \:\;\:\;\:  
\triangle_E = \frac{\sqrt{\displaystyle\sum_i c_i^2m_i^4}}{M^2}{\rm .}
\end{equation}
To apply our tuning measure in the three dimensional case we must
determine volumes $F$ and $G$. For a point, $(m_1^2, m_2^2, m_3^2)$,
with $M^2= M_{0}^2 = c_1 m_1^2 - c_2 m_2^2 + c_3 m_3^2$ we have,
\begin{eqnarray}\label{diff_F}
\frac{\partial^3 F}{\partial p_1^2 \partial p_2^2 \partial p_3^2} 
&=& \displaystyle\prod\limits_{i=1}^3\theta(p_i^2 - a m_i^2) 
\theta(b m_i^2 - p_i^2){\rm ,}\\
\label{diff_G}
\frac{\partial^3 G}{\partial p_1^2 \partial p_2^2 \partial p_3^2} 
&=& \frac{\partial^3 F}{\partial p_1^2 \partial p_2^2 \partial p_3^2}
\theta(M^2 - a M_{0}^2) \theta(b M_{0}^2 - M^2){\rm ,}
\end{eqnarray}
where the latter uses $M_i^2 = p_i^2$ and $\theta(x)$ is the usual
Heaviside step function. Integrating Eq.\ (\ref{diff_F}) over all
three $p_i$ gives the volume,
\begin{equation} F = (b-a)^3m_1^2m_2^2m_3^2 {\rm ,} \end{equation}
and similarly Eq.\ (\ref{diff_G}) gives, 
\begin{eqnarray} 
G& = &(b-a)^3 
\left\{ \theta(c_3 m_3^2 - c_2 m_2^2) \theta(c_2 m_2^2 - c_1 m_1^2)
\left[ \frac{1}{c_3}m_1^2 m_2^2M^2 - \frac{c_1^2}{3c_2 c_3}m_1^6 \right] 
\right. \nonumber \\ 
&+& \theta(c_1 m_1^2 - c_2 m_2^2) \theta(c_2 m_2^2 - c_3 m_3^2)
\left[ \frac{1}{c_1} m_2^2 m_3^2 M^2 - \frac{c_3^2}{3c_2 c_1}m_3^6 \right] 
\nonumber \\ 
&+& \theta(c_3 m_3^2 - c_2 m_2^2) \theta(c_1 m_1^2 - c_2 m_2^2)
\left[ m_1^2 m_2^2 m_3^2 - \frac{c_2^2}{3c_1 c_3}m_2^6 \right] \nonumber \\  
&+& \theta(c_2 m_2^2 - c_1 m_1^2) \theta(c_2 m_2^2 - c_3 m_3^2)
\left. \left[ \frac{1}{c_2}m_1^2 m_3^2M^2 - \frac{1}{3 c_1 c_2 c_3}M^6 \right] 
\right\} .
\end{eqnarray}   
We find that the analytical expressions for tuning in this model
depend on the mass hierarchy of $m_1$, $m_2$ and $m_3$. \\

\begin{figure}[ht]
\begin{center}
\includegraphics[height=50mm]{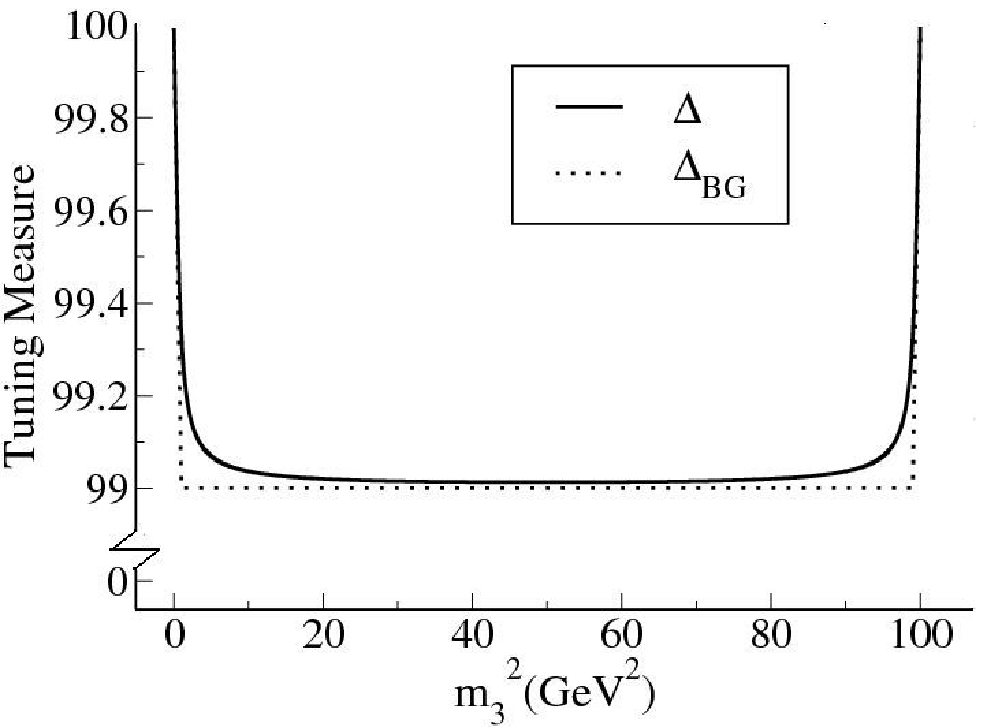} \hspace*{5mm}
\includegraphics[height=50mm]{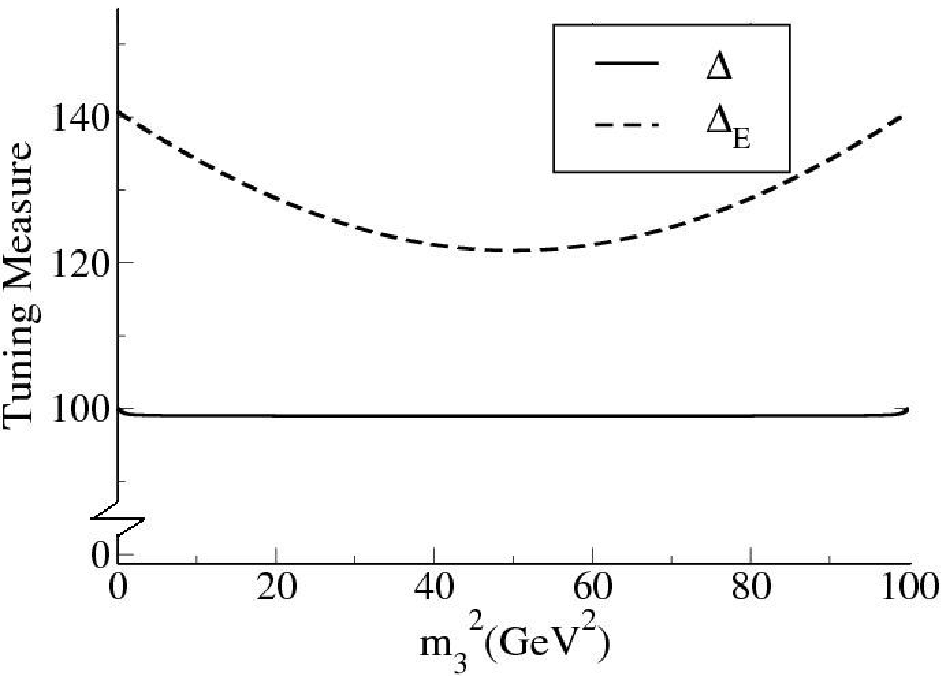}
\end{center}
\caption{Comparison of (unnormalised) tuning measures in the three
parameter model with $m_3^2$ varying from $0$ to $100
\,{\rm GeV}^2$ and $M_0^2 = 1 \,{\rm GeV}^2$ and $m_2^2= 99
\,{\rm GeV}^2$ kept constant. $m_1^2$ then varies according to Eq.\
(\ref{3d_tuning}) to accommodate the changes in $m_3^2$. {\it Left}:
between $\triangle_{BG}$ and our new measure. {\it Right}: between
$\triangle_E$ and our new measure.}
\label{var_m3_det_m1_BG_V_new}
\end{figure}

\noindent For $c_1 m_1^2 > c_2 m_2^2 > c_3 m_3^2$ we find,
\begin{eqnarray} 
\triangle =  \frac{F}{G} = 
\frac{c_1m_1^2 m_2^2 }{m_2^2 M^2 -\frac{c_3^2}{3c_2}m_3^4}
\approx \triangle_{BG} \;\;\;\; {\rm if} \;\;\;\;\; 
c_3m_3^2 \ll c_2 m_2^2. 
 \end{eqnarray} 
For $c_3 m_3^2 > c_2 m_2^2 > c_1 m_1^2$ we find:
\begin{eqnarray} 
\triangle = \frac{F}{G} = 
\frac{c_3m_2^2 m_3^2}{m_2^2 M^2 
-\frac{c_1^2}{3c_2}m_1^4}
\approx \triangle_{BG} 
\;\;\;\; {\rm if} \;\;\;\;\; 
c_1m_1^2 \ll c_2 m_2^2. 
\end{eqnarray}
For $c_3m_3^2 > c_1m_1^2 >c_2 m_2^2 $ and $ c_1m_1^2 > c_3m_3^2 >c_2m_2^2$:
\begin{eqnarray} 
\triangle = \frac{F}{G} 
= \frac{m_1^2 m_3^2}{m_1^2 m_3^2 -\frac{c_2^2}{3c_1 c_3}m_2^4}
\approx 1 \;\;\;\; {\rm if} \;\;\;\;\; c_1 m_1^2 c_3 m_3^2 \gg c_2^2 m_2^4. 
\end{eqnarray}
For $c_2m_2^2 > c_1m_1^2 >c_3 m_3^2 $ and $ c_2m_2^2 > c_3m_3^2 >c_1m_1^2$:
\begin{eqnarray} 
\triangle = \frac{F}{G} 
= \frac{c_2 m_1^2 m_2^2 m_3^2}{m_1^2 m_3^2 M^2 -\frac{1}{3 c_1 c_3}M^6}
\approx \triangle_{BG} \;\;\;\; {\rm if} \;\;\;\;\; 
M^4 \ll c_1 m_1^2 c_3 m_3^2.
\end{eqnarray}
Notice that these results do not match $\triangle_E$, but in three
dimensions at least $\triangle_{BG}$ is a much better
approximation, as is shown in
Fig.~\ref{var_m3_det_m1_BG_V_new}.

However, as we have seen, in moving from two parameters to three
parameters these discrepancies appeared, increasing the number of
parameters further will increase the divergences between the measures.

\section{Fine Tuning in the MSSM}
\label{MSSM_Tuning}
 The analytical methods described above become increasingly complicated
to apply as the number of parameters and observables are increased.  For
such situations we have also developed a numerical procedure which can
be applied to produce approximate results for tuning. Since the MSSM
contains many parameters and many observables we chose to apply our
numerical approach here. 

 We take random dimensionless fluctuations about an MSSM point at the
 GUT scale, $P' = \{p_k\}$, to give new points $\{P_i\}$. These are
 passed to a modified version of Softsusy
 2.0.5\cite{Allanach:2001kg}. Each random point $P_i$ is run down from
 the GUT scale until Electroweak Symmetry is broken. An iterative
 procedure is used to predict $M_Z^2$ and then all the sparticle and
 Higgs masses are determined. For a theoretical discussion, see
 Ref.\cite{Barger:1993gh}.

As before $F$ is the volume formed by dimensionless variations in the
parameters. $G_{O_i}$ is the sub-volume of $F$ additionally restricted
by dimensionless variations in the single observable $O_i$,
$a\leq\frac{O_i(\{p_k\})}{O_i(\{p_k'\})}\leq b$. As usual $G$ is the
volume restricted by $a\leq\frac{O_j(\{p_k\})}{O_j(\{p_k'\})}\leq b$,
for each observable, $O_j$, where $\{O_j\}$ is the set of masses
predicted in Softsusy. For every $O_i$ a count, $N_{O_i}$, is kept of
how often the point lies in the volume $G_{O_i}$ as well as an overall
count, $N_O$, kept of how many points are in $G$. Tuning is then
measured according to,
\begin{equation}\triangle_{O_i} \approx
\frac{N}{N_{O_i}}{\rm ,}\end{equation} 
for individual observables and
\begin{equation}\triangle \approx \frac{N}{N_O}\end{equation} 
for the overall tuning at that point.  \\

Before describing the results two comments on this approach should be
made. Firstly when using Softsusy to predict the masses for the random
points, sometimes problems are encountered. We may have a tachyon, the
Higgs potential unbounded from below, or non-perturbativity. Such
points don't belong in volume $G$ as they will give dramatically
different physics. However it is unclear which volumes, $G_{O_i}$, the
point lies in. Such points never register as hits in any of the
$G_{O_i}$ and this may artificially inflate the individual tunings,
including $\triangle_{M_Z^2}$. Keeping the range small reduces the
number of problem points. Therefore we chose $a=0.9$ and $b=1.1$ for our
dimensionless variations.

Secondly, since we are measuring tuning for individual points
numerically and cover only a small sample of points, it is not
possible to obtain mean values of $\triangle$ and the
$\triangle_{O_i}$ as we haven't sampled the entire space. When
simply comparing how the tuning varies about the parameter space the
normalisation factor is not needed. However to compare the tuning
between different observables as well as to compare with different
models some form of normalisation is essential.
\\

We considered points on the Constrained Minimal Supersymmetric
Standard Model (CMSSM) benchmark slope\footnote{Such benchmark slopes
and points, known as Snowmass Points and slopes
(SPS)\cite{Allanach:2002nj} are chosen by consensus as representing
qualitatively different MSSM scenarios and are very useful for
comparison with other work. }, SPS 1a \cite{Allanach:2002nj}. This
slope is defined by,
\begin{equation}\label{sps1a_slope}
m_0 = -A_0 = 0.4m_{\frac{1}{2}},\;\;\; 
{\rm sign}(\mu) = + ,\;\;\;
\tan{\beta} = 10 {\rm ,}
\end{equation}
 where $m_0$ is the common scalar mass, $m_{1/2}$ the common
 gaugino mass (both at the GUT scale) and ${\rm sign}(\mu)$ is the
 undetermined sign of $\mu$, the magnitude being determined from a
 loop corrected, inverted form of Eq.\ (\ref{Mzsqpred}) with $M_Z^2$
 set to it's observed value. $A_0$ is the common multiplicative factor
 which relates the supersymmetry breaking matrices of trilinear mass
 couplings to their corresponding Yukawa matrix, e.g.~$a_u = A_0 y_u$.

The parameters we vary simultaneously are the set\footnote{Note that
since points on the SPS 1a slope have $|\mu|$ set by $M_Z^2$, our
tuning measure is not sensitive to the $\mu$-problem.  However for our
random variation about the SPS 1a points we do treat $\mu_{GUT}$ as a
parameter because we are predicting $M_Z^2$ from the parameters, not
fixing it to it's observed value. } $\{m_0, m_{1/2}, \mu_{GUT},
m_3^2, A_0, y_t, y_b, y_{\tau} \}$, where $m_3$ is the soft bilinear
Higgs mixing parameter and $y_t, y_b, y_{\tau}$ are the Yukawa
couplings of the top, bottom and tau respectively. The gauge couplings
are not included as parameters. Doing so would introduce excessive
global sensitivity, increasing the statistics needed to keep
the errors under control.

\begin{figure}[ht]
\begin{center}
    \includegraphics[height=70mm, clip=true]{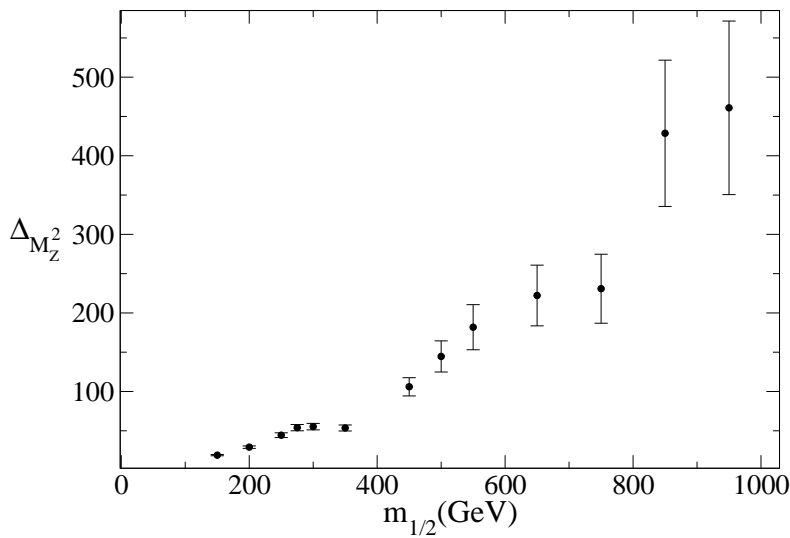}
\end{center}
\caption {$\triangle_{M_Z^2}$ for the SPS 1a slope. Error bars denote
a one standard deviation statistical error arising from the numerical
procedure. \label{fig:sps1amzt}}
\end{figure}

First we applied our tuning measure to the observable
$M_Z^2$ for 13 points on the SPS~1a slope. Moving along this slope
in $m_{1/2}$ is an increase in the overall supersymmetry breaking
scale, since the magnitude of every soft breaking term is
increasing. We have plotted the results of this investigation in Figure \ref{fig:sps1amzt}.

As expected there is a clear increase in tuning as the supersymmetry
breaking scale is raised. The statistical error also increases with
the tuning, making the numerical approach most difficult to apply when
the tuning is large. However precise determinations of tuning are only
relevant for moderate and low tunings. With tunings greater than
$500$, precise values are not required.
  
Due to the difficulty in this approach for measuring large tunings we
looked in more detail at points expected to have moderate tuning. We
chose a grid of points with,
\begin{eqnarray}A_0 = -100 \,{\rm GeV},\;\;\;\;\;\; 
\tan{\beta}= 10,\;\;\;\;\; {\rm sign}(\mu) = +, \nonumber \\  
250\,{\rm GeV} \leq m_{\frac{1}{2}} \leq 500\,{\rm GeV}{\rm ,} \;\;\;\:\:\: 
100\,{\rm GeV} \leq m_{0} \leq 200\,{\rm GeV}.
\end{eqnarray}

\begin{figure}[h!]
\begin{center}
\includegraphics[height=60mm,clip=true]{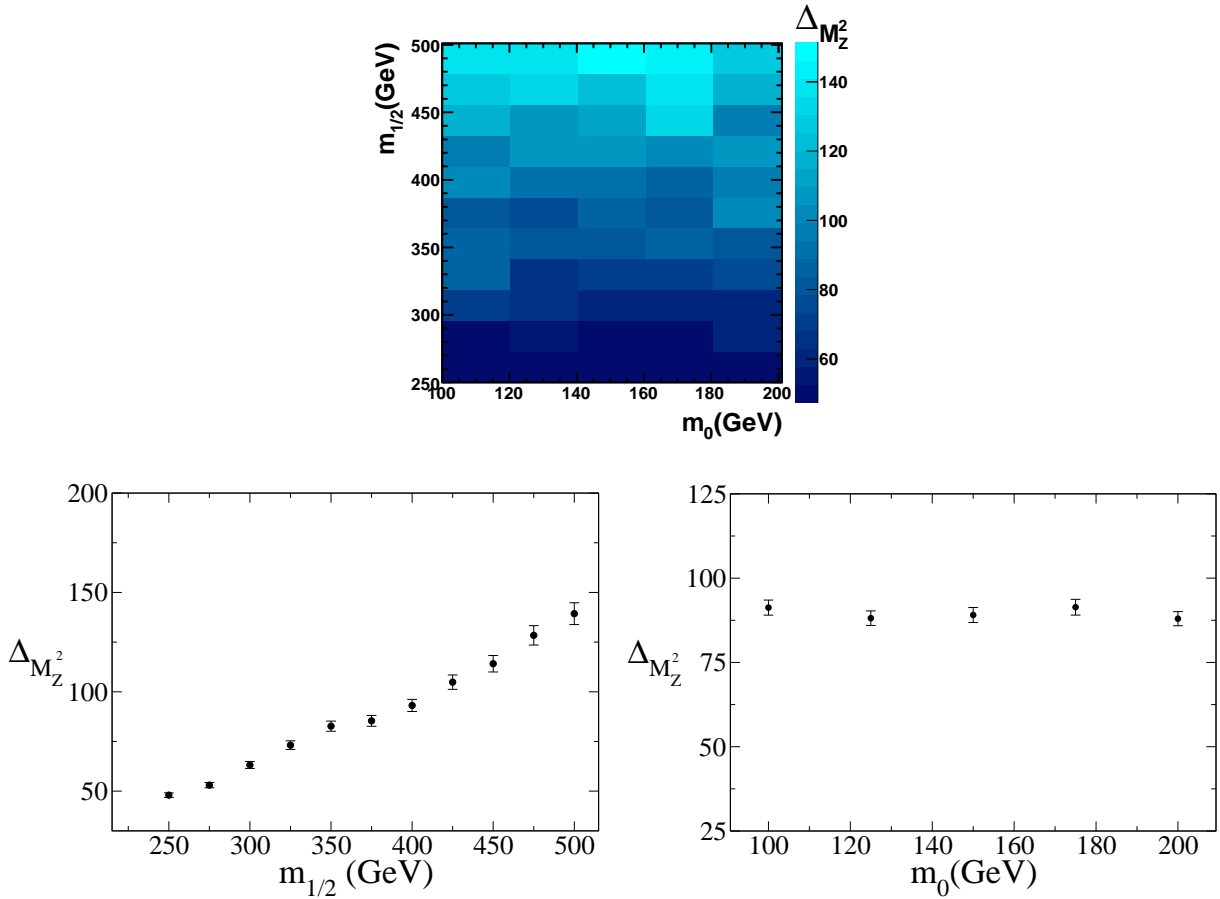} \\[4mm]
\includegraphics[height=55mm,clip=true]{mztuning_m12.eps} \hfill
\includegraphics[height=55mm,clip=true]{mztuning_m0.eps}
\end{center}
\caption{Tuning variation in $M_Z^2$. {\it Top}: $\triangle_{M_Z^2}$
for all points on our grid. {\it Bottom left}: $\triangle_{M_Z^2}$
plotted against $m_{1/2}$.  To reduce statistical errors, at each
value of $m_{1/2}$, we have taken the mean value $\triangle_{M_Z^2}$
over the five different $m_0$ values.  {\it Bottom right}:
$\triangle_{M_Z^2}$ plotted against $m_0$. To reduce statistical
errors, at each value of $m_0$, we have taken the mean value
$\triangle_{M_Z^2}$ over the eleven different $m_{1/2}$ values.}
\label{Tuning_Mz_figs}
\end{figure}

Shown in Fig.~\ref{Tuning_Mz_figs} (top) is a plot of
$\triangle_{M_Z^2}$ over this grid of points.  While the errors are
still significant ($\lesssim 10 \%$) there is a clear trend of tuning
increasing with $m_{1/2}$. Also shown (bottom left) is
$\triangle_{M_Z^2}$ averaged over the five different values of
$m_0$. This substantially reduces the errors giving a much more stable
picture of tuning increasing linearly with $m_{1/2}$.  Similarly
$\triangle_{M_Z^2}$, averaged over the eleven different values of
$m_{1/2}$, is shown (bottom right) as a function of
$m_0$. $\triangle_{M_Z^2}$ appears insensitive to variations in $m_0$.
These trends can be understood by looking at the one loop
renormalisation group improved version of Eq.\ (\ref{Mzsqpred}),
written in terms of the the parameters (with $\tan{\beta}= 10$),
\begin{equation}\label{one_loop_M_Z^2}
M_Z^2 \approx 2(-|\mu|^2 + 0.076 m_0^2 + 1.97 m_\frac{1}{2}^2 + 0.10 A_0^2 + 0.38 A_0 m_\frac{1}{2}){\rm ,}
\end{equation}
where $|\mu|^2$ is the value at $M_Z$ and and differs from the
parameter at the GUT scale, $\mu_{GUT}$. The large coefficient in
front of $m_{1/2}$ explains why explains why variations in this
parameter have a much greater impact on $\triangle_{M_Z^2}$ than
variations in $m_0$ whose coefficient is much smaller.

\begin{figure}[ht]
\begin{center}
\includegraphics[height=60mm,clip=true]{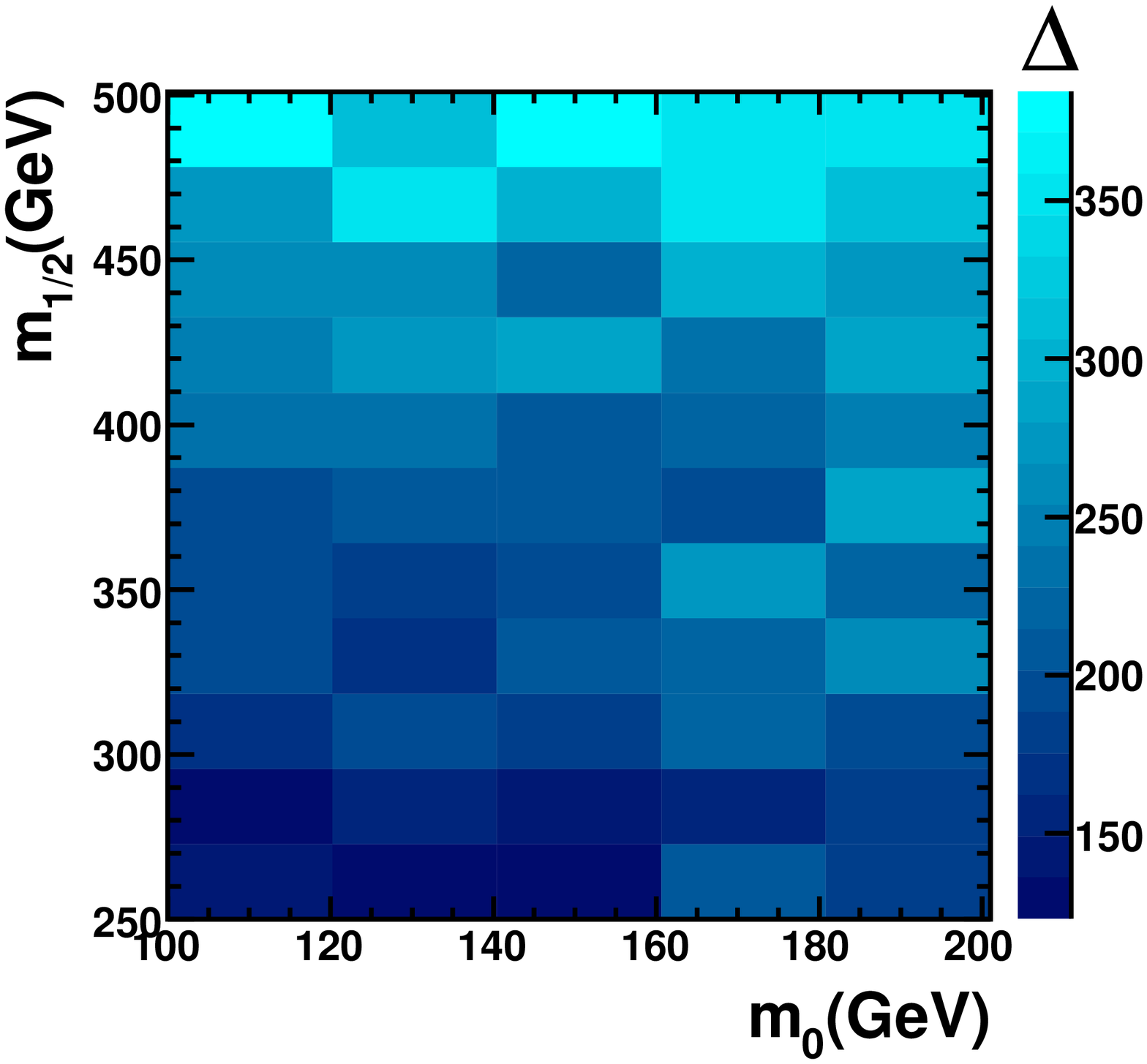} \\[4mm]
\includegraphics[height=55mm,clip=true]{alltuning_m12.eps} \hfill
\includegraphics[height=55mm,clip=true]{alltuning_m0.eps}
\end{center}
\caption{Variation in $\triangle$ plotted as in
Fig.~\ref{Tuning_Mz_figs} for $\triangle_{M_Z^2}$.}
\label{Tuning_all_figs}
\end{figure}

$\triangle$, which includes all of the masses predicted by Softsusy as
well as $M_Z^2$, is shown in Fig.~\ref{Tuning_all_figs}. Although the
errors are much larger here, a similar pattern to that for $M_Z^2$ can
be seen.  Since these are unnormalised tunings, the numerical values
of the two measures cannot be compared and one should not assume that
$\triangle > \triangle_{M_Z^2}$ implies that the tuning is worse than
when only $M_Z^2$ was considered.  In fact the lack of evidence for
distinct patterns of variation in tuning from the
Figs.~\ref{Tuning_Mz_figs} and \ref{Tuning_all_figs} is consistent with
the conjecture that the large cancellation between parameters in
$M_Z^2$ is the dominant source of the tuning for these points.

\begin{figure}[h!]
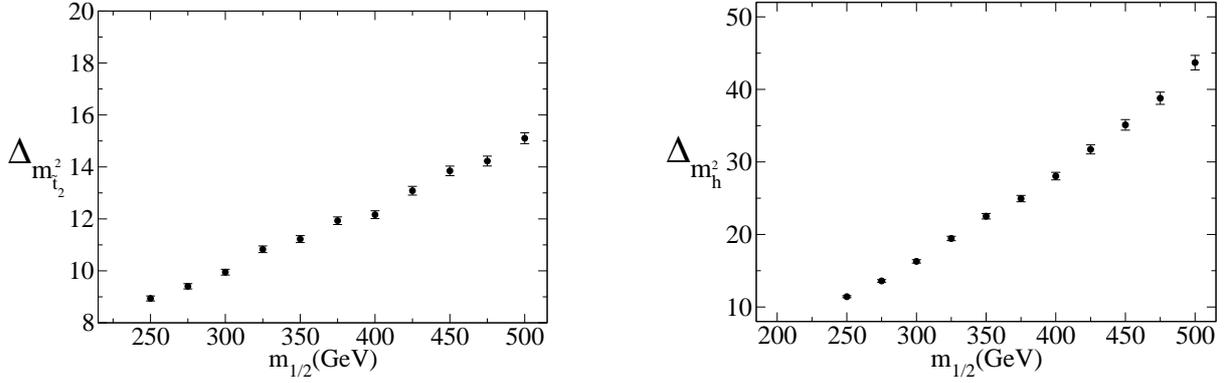

\begin{center}
\includegraphics[height=50mm,clip=true]{stoptuning_m12.eps} \hfill
\includegraphics[height=50mm,clip=true]{higgstuning_m12.eps}
\end{center}
\caption{Variation of unnormalised tunings in the mass of the heaviest
stop ($\triangle_{m^2_{\tilde{t}_2}}$, shown left) and the mass of the
light Higgs ($\triangle_{m_h^2}$, shown right) over $m_{1/2}$ }
\label{stop_higgs}
\end{figure}

Fig.~\ref{stop_higgs} shows that $\triangle_{m^2_{\tilde{t}_2}}$ and
$\triangle_{m^2_h}$ have similar patterns of variation to
$\triangle_{M_Z^2}$ and $\triangle$ over $m_{1/2}$, though the
gradients are noticeably shallower. While we know $m_h^2$ and
$m^2_{\tilde{t}_2}$ contribute to the Little Hierarchy Problem by
giving a large contribution to $M_Z^2$, thereby requiring a
cancellation to keep $M_Z$ light, this shows there is also some
tension in their own masses which restricts the parameter space. It is
not clear from our results whether or not dimensionless variations are
restricting different regions of parameter space to those in $M_Z^2$
or if $G_{m^2_{\tilde{t}_2}}$ and $G_{m^2_h}$ are merely sub-volumes
of $G_{M_Z^2}$, with no influence on $\triangle$. This topic deserves
further study.

\begin{figure}[h!]
\begin{center}
 \includegraphics[height=60mm,clip=true]{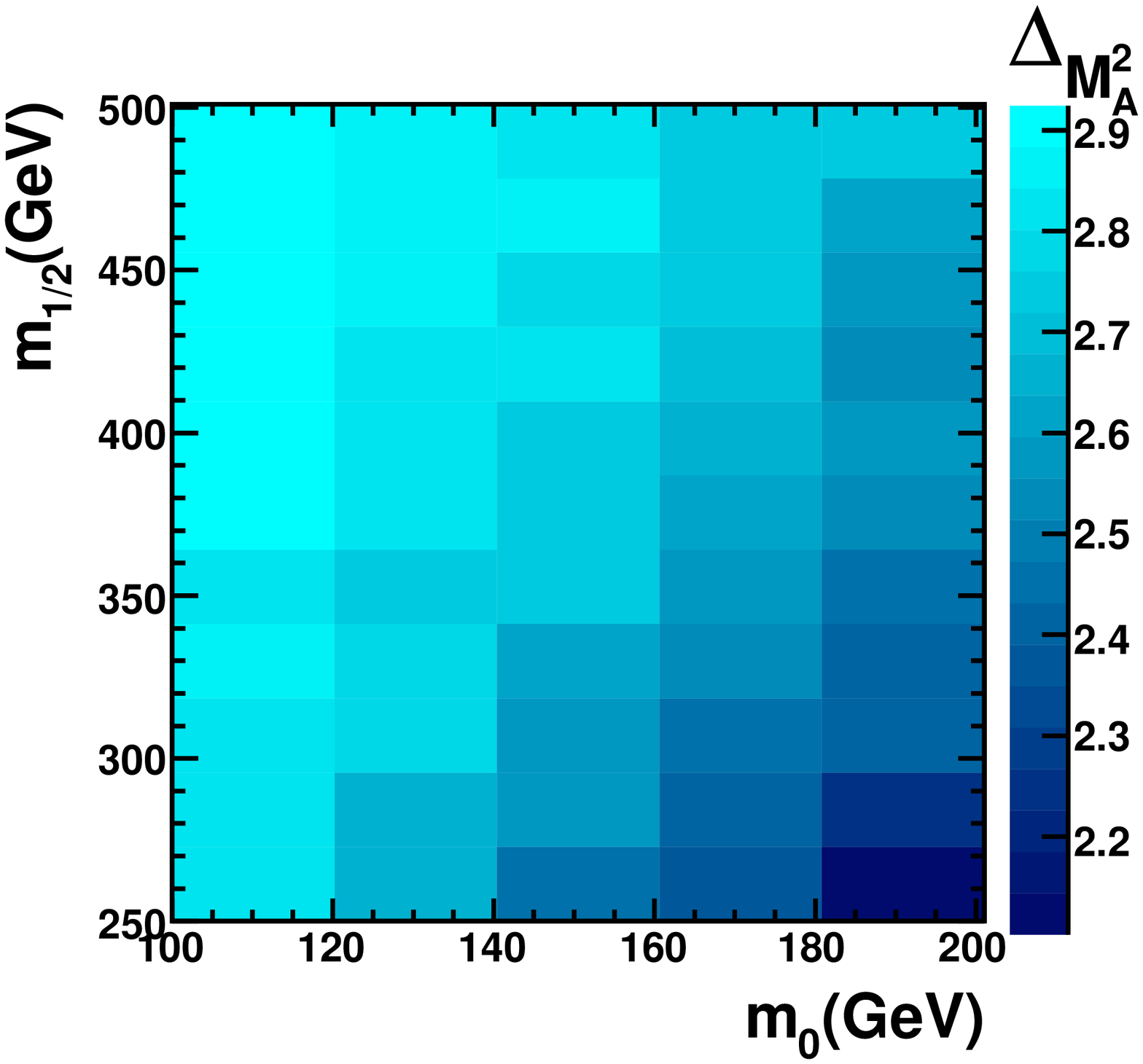}  \\[4mm]
 \includegraphics[height=53mm,clip=true]{matuning_m12.eps} \hfill
 \includegraphics[height=53mm,clip=true]{matuning_m0.eps}
\end{center}
\caption{Variation in $\triangle_{M_A^2}$ plotted as in
Fig.~\ref{Tuning_Mz_figs} for $\triangle_{M_Z^2}$.  }
\label{MA_tuning}
\end{figure}

However our results do show some evidence that the Little Hierarchy
Problem is not the only source of tuning. Displayed in
Fig.~\ref{MA_tuning} is $\triangle_{M_A^2}$.  Notice that
$\triangle_{M_A^2}$ is very small, so the errors are significantly
reduced and we can resolve very small variations in
$\triangle_{M_A^2}$. As with the other observables tuning increases
with $m_{1/2}$, but it is a distinctly non-linear variation. More
surprising is that tuning {\it decreases} with $m_0$. This pattern of
variation, distinct from that shown for $\triangle_{M_Z^2}$, shows a
different source of tension. It can be understood by examining the one
loop RGE solution for $M_A$,
\begin{equation}\label{M_A_eqn} 
M_A^2 \approx 2f(|\mu_{GUT}|^2,\{g_i\},\{y_i\}) + 0.81 m_0^2 - 1.55 m^2_\frac{1}{2} - 0.022 A_0^2 - 0.41 A_0 m_\frac{1}{2},
\end{equation}
where $f$ is a function of supersymmetry preserving parameters only,
arising from the evolution of $|\mu|^2$. Notice that there is some
opportunity for a cancellation here to make $M_A$ lighter than
expected. However the cancellation in the points we have looked at is
very small, leading to small values for $\triangle_{M_A^2}$. As
$m_0^2$ increases the already dominant positive part of the equation
increases and $M_A$ increases. As this happens the cancellation
becomes less significant to $M_A$ further reducing $\triangle_{M_A^2}$
as shown in Fig.~\ref{MA_tuning}(bottom right). Increasing $m_{1/2}$
increases the size of the cancellation. If all other parameters on the
right hand side of Eq.\ (\ref{M_A_eqn}) were fixed then we would
expect to see $\triangle_{M_A^2}$ increase linearly\footnote{The
effect of $A_0 m_{1/2}$ can be neglected since $m_{1/2}> A_0$.} with
$m_{1/2}$.  However each point on our grid has the value of $M_Z=91.188\,$GeV fixed, and the term $f(|\mu|^2,\{g_i\},\{y_i\}) \approx
|\mu^2|$ changes according to an inverted Eq.\ (\ref{one_loop_M_Z^2}).
This means $M_A^2$ is also increasing with $m_{1/2}$ and the balancing
act between these two different effects leads to the nonlinear pattern
shown.

\begin{figure}[ht]
\begin{center}
\includegraphics[height=70mm]{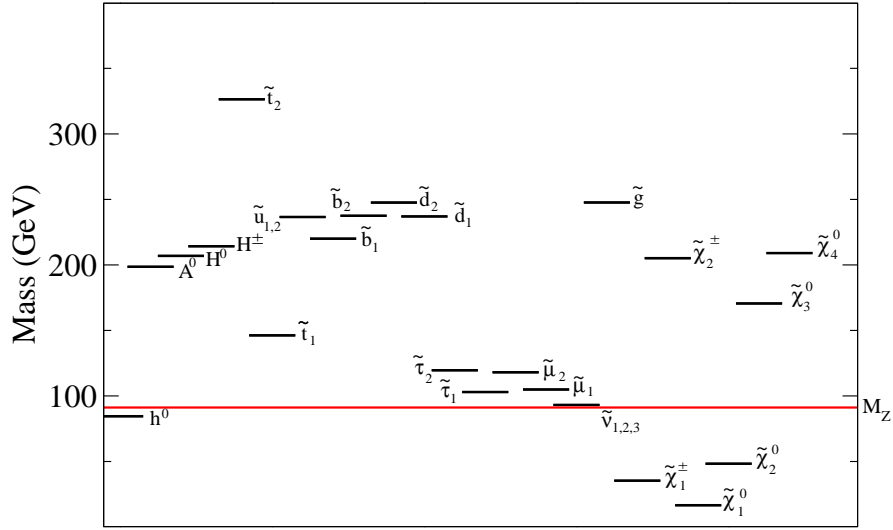}
\end{center}
\caption{Point NP1 with a ``natural'' spectrum}
\label{NP_1_spectrum}
\end{figure}

\begin{figure}[ht]
\begin{center}
\includegraphics[height=70mm]{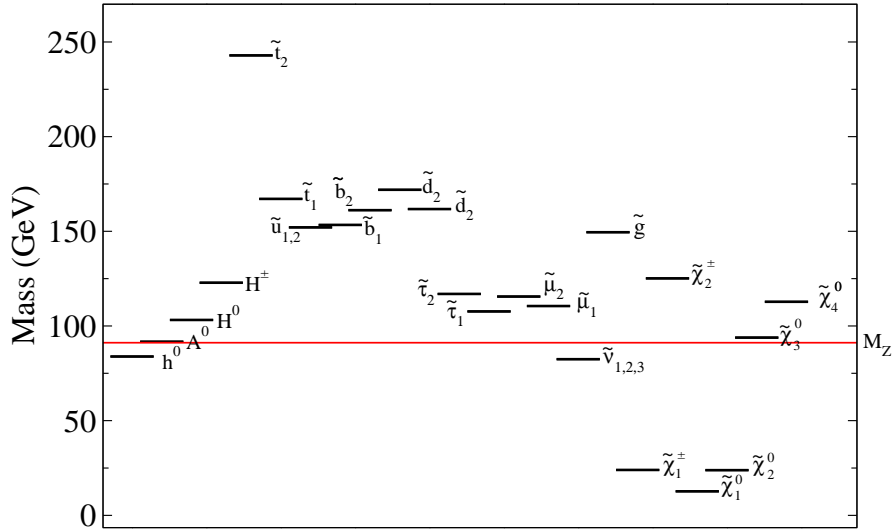}
\end{center}
\caption{Point NP2 with a ``natural'' spectrum}
\label{NP_2_spectrum}
\end{figure}

Although we can't determine the normalisation using this approach it
is nonetheless interesting to compare the unnormalised tunings for the
points in our study with those obtained for points with more
``natural'' looking spectra.  We present two points for this
purpose. NP1 and NP2 are defined by,
\begin{equation}
\begin{array}{lrclrclrclrclrcl}
\hspace*{-3mm} {\rm NP1:} 
& m_{\frac{1}{2}} & \hspace*{-3mm} = \hspace*{-3mm} & M_Z{\rm ,} 
& m_0 &\hspace*{-3mm} = \hspace*{-3mm}& M_Z{\rm ,}
& a_0 &\hspace*{-3mm} = \hspace*{-3mm}& -M_Z{\rm ,}
& {\rm sign}(\mu) &\hspace*{-3mm} = \hspace*{-3mm}& +{\rm ,}
& \tan{\beta} &\hspace*{-3mm} = \hspace*{-3mm}& 3, \\
\hspace*{-3mm} {\rm NP2:} 
& m_{\frac{1}{2}} & \hspace*{-3mm} = \hspace*{-3mm} & -50\, {\rm GeV,} 
& m_0 &\hspace*{-3mm} = \hspace*{-3mm}& 100\, {\rm GeV,} 
& a_0 &\hspace*{-3mm} = \hspace*{-3mm}& -50\, {\rm GeV,} 
& {\rm sign}(\mu) &\hspace*{-3mm} = \hspace*{-3mm}& +{\rm ,}
& \tan{\beta} &\hspace*{-3mm} = \hspace*{-3mm}& 10.
\end{array}
\end{equation}
 The spectra of these points are displayed in
Fig.~\ref{NP_1_spectrum} and Fig.~\ref{NP_2_spectrum}, and the
unnormalised tunings are displayed in Table
\ref{Tuning_comp_points}. Note that these are not intended to be
``realistic'' scenarios. Indeed both NP1 and NP2 are ruled out by
experiment but are simply intended to provide ``natural'' scenarios
for comparison.

\setlength{\tabcolsep}{4mm}
\begin{table}
\begin{center}
\begin{tabular}{l|cccccc} 
& $\triangle$ 
& $\triangle_{M_Z^2}$ 
& $\triangle_{m_{\tilde{t_2}}}$ 
& $\triangle_{m_{h}^2}$   
& $\triangle_{M_{A}^2}$  
& $\triangle_{m_{\chi_1^0}^2}$\\ \hline\\
{\rm NP1}  
& $241^{+36}_{-26}$ 
& $14.7^{+0.5}_{-0.5}$ 
& $6.7^{+0.1}_{-0.1}$ 
& $1.72^{+0.02}_{-0.02}$  
& $2.05^{+0.02}_{-0.02}$  
& $30.1^{+1.4}_{-1.3}$\\ \\
{\rm NP2}       
& $31.4^{+1.5}_{-1.4}$  
& $2.92^{+0.04}_{-0.04}$  
& $2.26^{+0.03}_{-0.03}$ 
& $1.87^{+0.02}_{-0.02}$ 
& $2.23^{+0.03}_{-0.03}$ 
& $2.64^{+0.04}_{-0.04}$\\  
\end{tabular}
\caption{Unnormalised tunings for the two points, NP1 and NP2, with
natural looking spectra.}
\label{Tuning_comp_points}
\end{center}
\end{table}

While NP1 has low values of $\triangle_{M_Z^2}$, $\triangle_{m_h^2}$,
$\triangle_{m^2_{\tilde{t}_2}}$ and $\triangle_{M^2_A}$, it has a
relatively large tuning in the mass of the lightest neutralino
($\triangle_{m_{\chi_1^0}^2}$). These combine to give a $\triangle$
which is similar in size to the values found for our grid of
points. In NP2 all of the tunings are relatively small, but the
combined tuning is still larger than may naively have been
anticipated. This is because many of these small tunings for individual
observables are not correlated and are restricting different regions
of parameter space. Table \ref{Relative_Tunings} shows the approximate
relative magnitude of the tunings in our grid points with respect to
these seemingly natural points.

\begin{table}
\begin{center}
\begin{tabular}{l|cccccc} 
& $\hat{\triangle}$ 
& $\hat{\triangle}_{M_Z^2}$ 
& $\hat{\triangle}_{m_{\tilde{t_2}}}$ 
& $\hat{\triangle}_{m_h^2}$   
& $\hat{\triangle}_{M_{A}^2}$  
& $\hat{\triangle}_{m_{\chi_1^0}^2}$\\ \hline\\
{\rm Relative to NP1}\,   
& $0.5..1.5$ 
& $3..10$ 
& $1..2$ 
& $7 ..25$  
& $ 1 $  
& $0.2 $\\ \\
{\rm Relative to NP2}\,   
& $5..15 $  
& $10..50$  
& $4..7$ 
& $ 6..23$ 
& $ 1 $ 
& $2$\\  
\end{tabular}
\caption{Approximate relative tunings for the points in our study,
with respect to those for NP1 and NP2.}
\label{Relative_Tunings}
\end{center}
\end{table}

In attempts to find a CMSSM scenario with a mass spectrum which is
manifestly natural we found many scenarios where tuning appeared in
the mass of the lightest neutralino. NP1 is a (moderate) example of
this.  This is because in some parameter choices, the lightest
neutralino becomes very light due to large cancellations between the
parameters. Other observables may also contain large cancellations
between the parameters in certain regions of parameter space. While we
have not studied this enough to make definitive claims, this may
suggest that mass hierarchies appear in a greater proportion of the
parameter space than conventional CMSSM wisdom dictates.  This would
reduce the true tuning in the CMSSM as scenarios with hierarchies
would be less atypical than previously thought.  A reduction in tuning
from this effect can only be measured by using our normalised new
measure, $\hat{\triangle}$.

Unfortunately the numerical approach we have applied to the MSSM in
this paper cannot be used to address this issue. An average measure of
$\triangle$, over the whole parameter space, is needed in order to
investigate this possibility.  A thorough numerical survey of the
parameter space would be too expensive, however an analytical study
may be more promising.  Findings in numerical studies like this may be
used to identify which observables and parameters are important for
fine tuning and therefore reduce the set $\{O_i\}$ and $\{p_i\}$ to a
manageable size. We will not carry out this programme here, but leave
it for a future study.

It is not just the possibility of finding a larger than expected
global sensitivity which motivates this study. It may be that most of
the CMSSM parameter space is hierarchy free and this is not a
significant effect.  However identifying a region of parameter space
where mass hierarchies are common also opens up new possibilities.
Past studies (see e.g.~\cite{Chankowski:1998xv,Kane:2002ap}) have
looked for a theoretical basis for relations between parameters which
enforce a hierarchy between $M_Z$ and $M_{SUSY}$. However no search
has been made for theoretical relations which simply restrict the
parameter space to regions where hierarchies, in general, are common.
Such studies may also have the possibility of solving the Little
Hierarchy Problem.

Here we have two complimentary approaches. An analytical approach
which can determine tuning precisely, but is complicated and unwieldy
when applied to a great number of parameters and observables and a
numerical approach which can be applied to such situations but is not
able to give an unambiguous measure of tuning as global sensitivity
cannot be accounted for. Progress can be made by combing our two
approaches.  Since solving for the tuning analytically with all
parameters and observables included would be difficult, one should
first apply the numerical method.  This might identify which
observables are in tension and responsible for the restriction of
parameter space and also along which axis in parameter space this
restriction takes place.  If these are a sufficiently small set (maybe
no more than 5 parameters and 5 observables) then the analytical
measure can be applied to this limited set to obtain a reasonably
accurate and unambiguous measure of tuning for that model.

\section{Conclusions}

Fine tuning $\approx 10^{34}$ within the Standard Model has motivated
many of the BSM theories which are popular within particle physics. In
particular it motivates low energy supersymmetry. However constraints
from LEP and other searches have placed stringent bounds on new
physics which mean that many of the proposed solutions to the SM fine
tuning problem also require tuning to some degree.  In order to
compare the viability of such models and judge whether or not they are
satisfactory a reliable measure of tuning is required.

Current measures of tuning have several limitations. They neglect the
many parameter nature of fine tuning, ignore additional tunings in
other observables, consider local stability only and assume ${\cal
L}_{SUSY}$ is parametrised in the same way as ${\cal L}_{GUT}$. In the
literature there have been different approaches to combine tunings for
individual parameters and observables. With no guiding principle to
select one particular approach, which models are preferred in terms of
naturalness can depend on which tuning measure is used.
 
In this paper we have presented a new measure of tuning based upon our
intuitive notion of the restriction of parameter space. This measure
can also be obtained by generalising the traditional measure of tuning
to include many parameters, many observables and finite variations in
the parameters followed by removing global sensitivity by factoring
out the mean value of the unnormalised sensitivity.

From the application of this new measure to various toy models, we
have shown that none of the other measures satisfactorily combine
individual tunings per parameter. Interestingly though, in the absence
of global sensitivity, it is the traditional measure of Barbieri and
Guidice which comes closest to our result with deviations for these
simple examples being very small.

A numerical approach for some CMSSM scenarios demonstrated how the
tuning in complicated models with many parameters and many observables
may be examined and also highlighted some of the complications and
issues encountered in doing so.

Our new measure is needed in future studies to examine tuning in the
$Z$ boson mass and cosmological relic density simultaneously; to judge
the true tuning in the NMSSM in light of \cite{Schuster:2005py}; to
examine parametrisation choices which alleviate the tuning in
different models and to study the global sensitivity of the complete
tuning measure to see if this may cause a significant reduction in the
tuning problem.


\begin{thebibliography}{9}

%\cite{Martin:1997ns}
\bibitem{Martin:1997ns}
  S.~P.~Martin,
  %``A supersymmetry primer,''
  arXiv:hep-ph/9709356.
  %%CITATION = HEP-PH/9709356;%%

%\cite{Yao:2006px}
\bibitem{Yao:2006px}
  W.~M.~Yao {\it et al.}  [Particle Data Group],
  %``Review of particle physics,''
  J.\ Phys.\ G {\bf 33} (2006) 1.
  %%CITATION = JPHGB,G33,1;%%

%\cite{Chang:2005ht}
\bibitem{Chang:2005ht}
  S.~Chang, P.~J.~Fox and N.~Weiner,
  %``Naturalness and Higgs decays in the MSSM with a singlet,''
  JHEP {\bf 0608}, 068 (2006)
  [arXiv:hep-ph/0511250].
  %%CITATION = JHEPA,0608,068;%%

\bibitem{Choi:2005uz}
  K.~Choi, K.~S.~Jeong and K.~i.~Okumura,
  %``Phenomenology of mixed modulus-anomaly mediation in fluxed string
  %compactifications and brane models,''
  JHEP {\bf 0509} (2005) 039
  [arXiv:hep-ph/0504037].
  %%CITATION = JHEPA,0509,039;%%

%\cite{Nomura:2005qg}
\bibitem{Nomura:2005qg}
  Y.~Nomura and B.~Tweedie,
  %``The supersymmetric fine-tuning problem and TeV-scale exotic scalars,''
  Phys.\ Rev.\  D {\bf 72}, 015006 (2005)
  [arXiv:hep-ph/0504246].
  %%CITATION = PHRVA,D72,015006;%%

\bibitem{Choi:2005hd}
  K.~Choi, K.~S.~Jeong, T.~Kobayashi and K.~i.~Okumura,
  %``Little SUSY hierarchy in mixed modulus-anomaly mediation,''
  Phys.\ Lett.\  B {\bf 633} (2006) 355
  [arXiv:hep-ph/0508029].
  %%CITATION = PHLTA,B633,355;%%

%\cite{Kitano:2005wc}
\bibitem{Kitano:2005wc}
  R.~Kitano and Y.~Nomura,
  %``A solution to the supersymmetric fine-tuning problem within the MSSM,''
  Phys.\ Lett.\  B {\bf 631}, 58 (2005)
  [arXiv:hep-ph/0509039].
  %%CITATION = PHLTA,B631,58;%%

%\cite{Lebedev:2005ge}
\bibitem{Lebedev:2005ge}
  O.~Lebedev, H.~P.~Nilles and M.~Ratz,
  %``A note on fine-tuning in mirage mediation,''
  arXiv:hep-ph/0511320.
  %%CITATION = HEP-PH/0511320;%%

\bibitem{Kitano:2006gv}
  R.~Kitano and Y.~Nomura,
  %``Supersymmetry, naturalness, and signatures at the LHC,''
  Phys.\ Rev.\  D {\bf 73}, 095004 (2006)
  [arXiv:hep-ph/0602096].
  %%CITATION = PHRVA,D73,095004;%%

%\cite{Chang:2006ra}
\bibitem{Chang:2006ra}
  S.~Chang, L.~J.~Hall and N.~Weiner,
  %``A supersymmetric twin Higgs,''
  Phys.\ Rev.\  D {\bf 75}, 035009 (2007)
  [arXiv:hep-ph/0604076].
  %%CITATION = PHRVA,D75,035009;%%

%\cite{Casas:2005ev}
\bibitem{Casas:2005ev}
  J.~A.~Casas, J.~R.~Espinosa and I.~Hidalgo,
  %``Implications for new physics from fine-tuning arguments. II: Little  Higgs
  %models,''
  JHEP {\bf 0503}, 038 (2005)
  [arXiv:hep-ph/0502066].
  %%CITATION = JHEPA,0503,038;%%

%\cite{Chacko:2005pe}
\bibitem{Chacko:2005pe}
  Z.~Chacko, H.~S.~Goh and R.~Harnik,
  %``The twin Higgs: Natural electroweak breaking from mirror symmetry,''
  Phys.\ Rev.\ Lett.\  {\bf 96}, 231802 (2006)
  [arXiv:hep-ph/0506256].
  %%CITATION = PRLTA,96,231802;%%

%\cite{Chacko:2005un}
\bibitem{Chacko:2005un}
  Z.~Chacko, H.~S.~Goh and R.~Harnik,
  %``A twin Higgs model from left-right symmetry,''
  JHEP {\bf 0601}, 108 (2006)
  [arXiv:hep-ph/0512088].
  %%CITATION = JHEPA,0601,108;%%

%\cite{Barbieri:1987fn}
\bibitem{Barbieri:1987fn}
  R.~Barbieri and G.~F.~Giudice,
  %``Upper Bounds On Supersymmetric Particle Masses,''
  Nucl.\ Phys.\  B {\bf 306}, 63 (1988).
  %%CITATION = NUPHA,B306,63;%%

\bibitem{Ellis:1986yg}
  J.~R.~Ellis, K.~Enqvist, D.~V.~Nanopoulos and F.~Zwirner,
  %``Observables In Low-Energy Superstring Models,''
  Mod.\ Phys.\ Lett.\  A {\bf 1} (1986) 57.
  %%CITATION = MPLAE,A1,57;%%

%\cite{deCarlos:1993yy}
\bibitem{deCarlos:1993yy}
  B.~de Carlos and J.~A.~Casas,
  %``One loop analysis of the electroweak breaking in supersymmetric models and
  %the fine tuning problem,''
  Phys.\ Lett.\  B {\bf 309}, 320 (1993)
  [arXiv:hep-ph/9303291].
  %%CITATION = PHLTA,B309,320;%%

%\cite{deCarlos:1993ca}
\bibitem{deCarlos:1993ca}
  B.~de Carlos and J.~A.~Casas,
  %``The Fine tuning problem of the electroweak symmetry breaking mechanism in
  %minimal SUSY models,''
  arXiv:hep-ph/9310232.
  %%CITATION = HEP-PH/9310232;%%

%\cite{Chankowski:1997zh}
\bibitem{Chankowski:1997zh}
  P.~H.~Chankowski, J.~R.~Ellis and S.~Pokorski,
  %``The fine-tuning price of LEP,''
  Phys.\ Lett.\  B {\bf 423}, 327 (1998)
  [arXiv:hep-ph/9712234].
  %%CITATION = PHLTA,B423,327;%%

%\cite{Agashe:1997kn}
\bibitem{Agashe:1997kn}
  K.~Agashe and M.~Graesser,
  %``Improving the fine tuning in models of low energy gauge mediated
  %supersymmetry breaking,''
  Nucl.\ Phys.\  B {\bf 507}, 3 (1997)
  [arXiv:hep-ph/9704206].
  %%CITATION = NUPHA,B507,3;%%

\bibitem{Wright:1998mk}
  D.~Wright,
  %``Naturally nonminimal supersymmetry,''
  arXiv:hep-ph/9801449.
  %%CITATION = HEP-PH/9801449;%%

\bibitem{Kane:1998im}
  G.~L.~Kane and S.~F.~King,
  %``Naturalness implications of LEP results,''
  Phys.\ Lett.\  B {\bf 451} (1999) 113
  [arXiv:hep-ph/9810374].
  %%CITATION = PHLTA,B451,113;%%
 
%\cite{Bastero-Gil:1999gu}
\bibitem{Bastero-Gil:1999gu}
  M.~Bastero-Gil, G.~L.~Kane and S.~F.~King,
  %``Fine-tuning constraints on supergravity models,''
  Phys.\ Lett.\  B {\bf 474}, 103 (2000)
  [arXiv:hep-ph/9910506].
  %%CITATION = PHLTA,B474,103;%%

\bibitem{Feng:1999zg}
  J.~L.~Feng, K.~T.~Matchev and T.~Moroi,
  %``Focus points and naturalness in supersymmetry,''
  Phys.\ Rev.\  D {\bf 61}, 075005 (2000)
  [arXiv:hep-ph/9909334].
  %%CITATION = PHRVA,D61,075005;%%

%\cite{Allanach:2000ii}
\bibitem{Allanach:2000ii}
  B.~C.~Allanach, J.~P.~J.~Hetherington, M.~A.~Parker and B.~R.~Webber,
  %``Naturalness reach of the Large Hadron Collider in minimal supergravity,''
  JHEP {\bf 0008}, 017 (2000)
  [arXiv:hep-ph/0005186].
  %%CITATION = JHEPA,0008,017;%%

%\cite{Allanach:2006jc}
\bibitem{Allanach:2006jc}
  B.~C.~Allanach,
  %``Naturalness priors and fits to the constrained minimal supersymmetric
  %standard model,''
  Phys.\ Lett.\  B {\bf 635}, 123 (2006)
  [arXiv:hep-ph/0601089].
  %%CITATION = PHLTA,B635,123;%%

%\cite{Kobayashi:2006fh}
\bibitem{Kobayashi:2006fh}
  T.~Kobayashi, H.~Terao and A.~Tsuchiya,
  %``Fine-tuning in gauge mediated supersymmetry breaking models and induced
  %top Yukawa coupling,''
  Phys.\ Rev.\  D {\bf 74}, 015002 (2006)
  [arXiv:hep-ph/0604091].
  %%CITATION = PHRVA,D74,015002;%%

%\cite{Dermisek:2005ar}
\bibitem{Dermisek:2005ar}
  R.~Dermisek and J.~F.~Gunion,
  %``Escaping the large fine tuning and little hierarchy problems in the  next
  %to minimal supersymmetric model and h --> a a decays,''
  Phys.\ Rev.\ Lett.\  {\bf 95}, 041801 (2005)
  [arXiv:hep-ph/0502105].
  %%CITATION = PRLTA,95,041801;%%

%\cite{Barbieri:2005kf}
\bibitem{Barbieri:2005kf}
  R.~Barbieri and L.~J.~Hall,
  %``Improved naturalness and the two Higgs doublet model,''
  arXiv:hep-ph/0510243.
  %%CITATION = HEP-PH/0510243;%%

%\cite{Barbieri:2006dq}
\bibitem{Barbieri:2006dq}
  R.~Barbieri, L.~J.~Hall and V.~S.~Rychkov,
  %``Improved naturalness with a heavy Higgs: An alternative road to LHC
  %physics,''
  Phys.\ Rev.\  D {\bf 74}, 015007 (2006)
  [arXiv:hep-ph/0603188].
  %%CITATION = PHRVA,D74,015007;%%

%\cite{Gripaios:2006nn}
\bibitem{Gripaios:2006nn}
  B.~Gripaios and S.~M.~West,
  %``Improved Higgs naturalness with or without supersymmetry,''
  Phys.\ Rev.\  D {\bf 74}, 075002 (2006)
  [arXiv:hep-ph/0603229].
  %%CITATION = PHRVA,D74,075002;%%

\bibitem{Dermisek:2006py}
  R.~Dermisek, J.~F.~Gunion and B.~McElrath,
  %``Probing NMSSM scenarios with minimal fine-tuning by searching for decays of
  %the Upsilon to a light CP-odd Higgs boson,''
  Phys.\ Rev.\  D {\bf 76} (2007) 051105
  [arXiv:hep-ph/0612031].
  %%CITATION = PHRVA,D76,051105;%%

%\cite{Anderson:1994dz}
\bibitem{Anderson:1994dz}
  G.~W.~Anderson and D.~J.~Castano,
  %``Measures of fine tuning,''
  Phys.\ Lett.\  B {\bf 347}, 300 (1995)
  [arXiv:hep-ph/9409419].
  %%CITATION = PHLTA,B347,300;%%

%\cite{Anderson:1994tr}
\bibitem{Anderson:1994tr}
  G.~W.~Anderson and D.~J.~Castano,
  %``Naturalness And Superpartner Masses Or When To Give Up On Weak Scale
  %Supersymmetry,''
  Phys.\ Rev.\  D {\bf 52}, 1693 (1995)
  [arXiv:hep-ph/9412322].
  %%CITATION = PHRVA,D52,1693;%%

%\cite{Anderson:1995cp}
\bibitem{Anderson:1995cp}
  G.~W.~Anderson and D.~J.~Castano,
  %``Challenging weak scale supersymmetry at colliders,''
  Phys.\ Rev.\  D {\bf 53}, 2403 (1996)
  [arXiv:hep-ph/9509212].
  %%CITATION = PHRVA,D53,2403;%%

%\cite{Anderson:1996ew}
\bibitem{Anderson:1996ew}
  G.~W.~Anderson, D.~J.~Castano and A.~Riotto,
  %``Naturalness lowers the upper bound on the lightest Higgs boson mass in
  %supersymmetry,''
  Phys.\ Rev.\  D {\bf 55}, 2950 (1997)
  [arXiv:hep-ph/9609463].
  %%CITATION = PHRVA,D55,2950;%%

%\cite{Casas:2003jx}
\bibitem{Casas:2003jx}
  J.~A.~Casas, J.~R.~Espinosa and I.~Hidalgo,
  %``The MSSM fine tuning problem: A way out,''
  JHEP {\bf 0401}, 008 (2004)
  [arXiv:hep-ph/0310137].
  %%CITATION = JHEPA,0401,008;%%

%\cite{Casas:2004uu}
\bibitem{Casas:2004uu}
  J.~A.~Casas, J.~R.~Espinosa and I.~Hidalgo,
  %``A relief to the supersymmetric fine tuning problem,''
  arXiv:hep-ph/0402017.
  %%CITATION = HEP-PH/0402017;%%

%\cite{Casas:2004gh}
\bibitem{Casas:2004gh}
  J.~A.~Casas, J.~R.~Espinosa and I.~Hidalgo,
  %``Implications for new physics from fine-tuning arguments. I: Application  to
  %SUSY and seesaw cases,''
  JHEP {\bf 0411}, 057 (2004)
  [arXiv:hep-ph/0410298].
  %%CITATION = JHEPA,0411,057;%%

%\cite{Casas:2006bd}
\bibitem{Casas:2006bd}
  J.~A.~Casas, J.~R.~Espinosa and I.~Hidalgo,
  %``Expectations for LHC from naturalness: Modified vs. SM Higgs sector,''
  Nucl.\ Phys.\  B {\bf 777} (2007) 226
  [arXiv:hep-ph/0607279].
  %%CITATION = NUPHA,B777,226;%%

%\cite{Ciafaloni:1996zh}
\bibitem{Ciafaloni:1996zh}
  P.~Ciafaloni and A.~Strumia,
  %``Naturalness upper bounds on gauge mediated soft terms,''
  Nucl.\ Phys.\  B {\bf 494}, 41 (1997)
  [arXiv:hep-ph/9611204].
  %%CITATION = NUPHA,B494,41;%%

\bibitem{Chan:1997bi}
  K.~L.~Chan, U.~Chattopadhyay and P.~Nath,
  %``Naturalness, weak scale supersymmetry and the prospect for the  observation
  %of supersymmetry at the Tevatron and at the LHC,''
  Phys.\ Rev.\  D {\bf 58}, 096004 (1998)
  [arXiv:hep-ph/9710473].
  %%CITATION = PHRVA,D58,096004;%%

%\cite{Barbieri:1998uv}
\bibitem{Barbieri:1998uv}
  R.~Barbieri and A.~Strumia,
  %``About the fine-tuning price of LEP,''
  Phys.\ Lett.\  B {\bf 433}, 63 (1998)
  [arXiv:hep-ph/9801353].
  %%CITATION = PHLTA,B433,63;%%

%\cite{Giusti:1998gz}
\bibitem{Giusti:1998gz}
  L.~Giusti, A.~Romanino and A.~Strumia,
  %``Natural ranges of supersymmetric signals,''
  Nucl.\ Phys.\  B {\bf 550}, 3 (1999)
  [arXiv:hep-ph/9811386].
  %%CITATION = NUPHA,B550,3;%%

%\cite{Chankowski:1998za}
\bibitem{Chankowski:1998za}
  P.~H.~Chankowski, J.~R.~Ellis, K.~A.~Olive and S.~Pokorski,
  %``Cosmological fine tuning, supersymmetry, and the gauge hierarchy
  %problem,''
  Phys.\ Lett.\  B {\bf 452}, 28 (1999)
  [arXiv:hep-ph/9811284].
  %%CITATION = PHLTA,B452,28;%%

%\cite{Ellis:2002rp}
\bibitem{Ellis:2002rp}
  J.~R.~Ellis, K.~A.~Olive and Y.~Santoso,
  %``Constraining supersymmetry,''
  New J.\ Phys.\  {\bf 4}, 32 (2002)
  [arXiv:hep-ph/0202110].
  %%CITATION = NJOPF,4,32;%%

\bibitem{King:2006tf}
  S.~F.~King and J.~P.~Roberts,
  %``Natural implementation of neutralino dark matter,''
  JHEP {\bf 0609} (2006) 036
  [arXiv:hep-ph/0603095].
  %%CITATION = JHEPA,0609,036;%%

\bibitem{King:2006cu}
  S.~F.~King and J.~P.~Roberts,
  %``Natural dark matter from type I string theory,''
  JHEP {\bf 0701} (2007) 024
  [arXiv:hep-ph/0608135].
  %%CITATION = JHEPA,0701,024;%%

%\cite{Allanach:2001kg}
\bibitem{Allanach:2001kg}
  B.~C.~Allanach,
  %``SOFTSUSY: A C++ program for calculating supersymmetric spectra,''
  Comput.\ Phys.\ Commun.\  {\bf 143}, 305 (2002)
  [arXiv:hep-ph/0104145].
  %%CITATION = CPHCB,143,305;%%

%\cite{Barger:1993gh}
\bibitem{Barger:1993gh}
  V.~D.~Barger, M.~S.~Berger and P.~Ohmann,
  %``The Supersymmetric particle spectrum,''
  Phys.\ Rev.\  D {\bf 49}, 4908 (1994)
  [arXiv:hep-ph/9311269].
  %%CITATION = PHRVA,D49,4908;%%

%\cite{Allanach:2002nj}
\bibitem{Allanach:2002nj}
  B.~C.~Allanach {\it et al.}, Eur. Phys. J. C {\bf 25} (2002) 113 
  [arXiv:hep-ph/0202233].

%\cite{Chankowski:1998xv}
\bibitem{Chankowski:1998xv}
  P.~H.~Chankowski, J.~R.~Ellis, M.~Olechowski and S.~Pokorski,
  %``Haggling over the fine-tuning price of LEP,''
  Nucl.\ Phys.\  B {\bf 544}, 39 (1999)
  [arXiv:hep-ph/9808275].
  %%CITATION = NUPHA,B544,39;%%

%\cite{Kane:2002ap}
\bibitem{Kane:2002ap}
  G.~L.~Kane, J.~D.~Lykken, B.~D.~Nelson and L.~T.~Wang,
  %``Re-examination of electroweak symmetry breaking in supersymmetry and
  %implications for light superpartners,''
  Phys.\ Lett.\  B {\bf 551}, 146 (2003)
  [arXiv:hep-ph/0207168].
  %%CITATION = PHLTA,B551,146;%%

%\cite{Schuster:2005py}
\bibitem{Schuster:2005py}
  P.~C.~Schuster and N.~Toro,
  %``Persistent fine-tuning in supersymmetry and the NMSSM,''
  arXiv:hep-ph/0512189.
  %%CITATION = HEP-PH/0512189;%%

\end{thebibliography}
\end{document}